\documentclass[letterpaper,twocolumn,10pt]{article}
\usepackage{usenix2019_v3,epsfig,endnotes}
\usepackage{booktabs}
\usepackage{times}
\usepackage{enumitem}
\usepackage{subfigure}
\usepackage{amsmath,graphicx,hyperref,listings,color,xspace}
\usepackage{datetime}
\usepackage{url}
\usepackage{textcomp}
\usepackage{comment}
\usepackage{xcolor}
\usepackage{colortbl}
\usepackage{multirow}
\usepackage{array}
\newcolumntype{P}[1]{>{\centering\arraybackslash}p{#1}}
\usepackage[normalem]{ulem}
\usepackage{kotex}
\usepackage{amsmath}

\usepackage{graphicx}
\usepackage{soul}
\usepackage{hyperref}
\usepackage{tikz}
\usepackage{afterpage}
\usepackage{adjustbox}
\usepackage{tabularx}
\usepackage{threeparttable}
\usepackage{amssymb}

\hypersetup{
    colorlinks=true,
    linkcolor=blue,
    filecolor=magenta,
    citecolor=blue,
    urlcolor=blue,
}

\newcommand{\SN}[1]{\textcolor{red}{#1}} 

\newcommand{\sys}{{RALP}}

\newboolean{showcomments}
\setboolean{showcomments}{true}

\newcommand{\mj}[1]{  \ifthenelse{\boolean{showcomments}}
{ \textcolor{blue}{(MJ:  #1)}} {}  }
\newcommand{\noh}[1]{  \ifthenelse{\boolean{showcomments}}
{ \textcolor{Blue}{(noh:  #1)}} {}  }
\newcommand{\jay}[1]{  \ifthenelse{\boolean{showcomments}}
{ \textcolor{purple}{(jay:  #1)}} {}  }
\newcommand{\sh}[1]{  \ifthenelse{\boolean{showcomments}}
{ \textcolor{brown}{(sh:  #1)}} {}  }

\newcommand{\Paragraph}[1]{\vskip 3pt\noindent\textbf{#1 }}	 

\makeatletter
\patchcmd{\@verbatim}
  {\verbatim@font}
  {\verbatim@font\footnotesize}
  {}{}
\makeatother

\setlist[itemize]{noitemsep, topsep=0pt}
\setlist[enumerate]{noitemsep, topsep=0pt}

\makeatletter
\patchcmd{\ttlh@hang}{\parindent\z@}{\parindent\z@\leavevmode}{}{}
\patchcmd{\ttlh@hang}{\noindent}{}{}{}
\makeatother

\begin{document}

\date{}

\title{\Large \bf Accelerated Training for CNN Distributed Deep Learning\\through Automatic Resource-Aware Layer Placement}

\author{
Jay H. Park, Sunghwan Kim, Jinwon Lee, Myeongjae Jeon, and Sam H. Noh\\
UNIST}

\maketitle


\begin{abstract}
The Convolutional Neural Network (CNN) model, often used for image
classification, requires significant training time to obtain high accuracy.
To this end, distributed training is performed with the parameter server
(PS) architecture using multiple servers. Unfortunately, scalability has
been found to be poor in existing architectures. We find that the PS
network is the bottleneck as it communicates a large number of gradients
and parameters with the many workers. This is because synchronization with
the many workers has to occur at every step of training. Depending on the
model, communication can be in the several hundred MBs per synchronization.
In this paper, we propose a scheme to reduce network traffic through layer
placement that considers the resources that each layer uses. Through
analysis of the characteristics of CNN, we find that placement of layers
can be done in an effective manner. We then incorporate this observation
within the TensorFlow framework such that layers can be automatically
placed for more efficient training. Our evaluation making use of this
placement scheme show that training time can be significantly reduced
without loss of accuracy for many CNN models.

\end{abstract}

\section{Introduction}
\label{sec:intro}

In recent years, machine learning has been used to solve problems in many
fields~\cite{Le2012highlevel, 2016lstm, zhang2016augmenting,
NIPS2016_6186,inception-v3_arxiv}. In particular, Convolutional Neural
Networks (CNNs) has been popular in vision and audio recognition
areas~\cite{2012alexnet, 2015vgg, 2016resnet}.
Various machine learning models go through a training phase, which makes use
of training data, that requires many iterative computations through the
model. To achieve high prediction accuracy through training, large scale
machine learning models such as deep learning are
used~\cite{chilimbi2014project}. Training using a large scale model requires
a significant amount of data and computation. This is generally achieved
through distributed execution~\cite{dean2012distbelief, xing2015petuum,
li2014scaling, chilimbi2014project} or by making use of
GPUs~\cite{awan2017scaffe, 2017poseidon, cui2016geeps}. To make use of large
scale executions, machine learning frameworks such as
TensorFlow~\cite{abadi2016tensorflow}, Caffe~\cite{jia2014caffe}, and
Torch~\cite{collobert2011torch7} are openly available.

Many modern distributed machine learning systems make use of the parameter
server
architecture~\cite{2017poseidon,cui2016geeps,dean2012distbelief,li2014scaling,xing2015petuum,harlap2017proteus}.
With this architecture, data parallelism and model parallelism is used to
compute a significant amount of training data~\cite{dean2012distbelief}.
For data parallelism, the model itself is replicated, while for model
parallelism, the model is partitioned on multiple GPUs and run in parallel.
Model parallelism can be useful for large models. However, it is rarely used
as the performance gains have been shown to be minimal due to heavy network
traffic~\cite{jia2018beyond}.
In this study, we focus on data parallelism in a distributed parameter server architecture.

Training in such an architecture, however, has its own limitations.
The first is network traffic. 
The network can become the bottleneck as the workers must
communicate and synchronize with the parameter server at each training step.
Such network traffic tend to grow with the number of parameter servers in
case multiple parameter servers are deployed for large parameter
sizes~\cite{peng2018optimus}. The second limitation comes with the issue of
placement. That is, models are placed at particular nodes such as servers
with GPUs and what not. Such placement decisions affect computation time as
well as network communication as these machines may need to transfer data
among themselves. Thus, ineffective placement can lead to heavy
communication, leading to degraded performance. Such placement decisions can
be difficult to make if one is not familiar with the characteristics of the
model~\cite{2017device_placement}.

The goal of this study is to improve performance in distributed deep learning training by alleviating the network bottleneck and balancing computation, with a focus on CNN.
We propose a Resource-Aware Layer Placement (RALP) scheme that minimizes network traffic, while considering the
computation overhead incurred by the placement. To develop this scheme, we go
through a careful analysis of various CNN models, where we characterize the
computational and communication needs of the layers that comprise CNN. We
find common traits that may be exploited in placing the layers. 
In particular, we find the fully connected layer is the best fit to be placed in
the parameter server. 
We implement RALP in the TensorFlow framework such that
the nodes comprising particular layers may be placed in appropriate servers
for improved performance. Our experimental results show that, while not all
models can benefit from RALP, many models are able to reap significant
performance improvements, improving by as much as 13$\times$ compared to the
baseline TensorFlow framework.

The rest of the paper is organized as follows. 
In the next section, we discuss CNN as well as distributed deep learning training as background.
In Section~\ref{section:Characterization}, we discuss the motivation behind our study, focusing on detailed characterizations and analysis of various components of CNN that affect performance. 
Based on this analysis, we present the design of Resource-Aware Layer Placement (RALP), the scheme that we propose in  Section~\ref{section:RALP} and then, its implementation issues in Section~\ref{section:impl}. 
We present experimental results that show that RALP is effective in Section~\ref{section:evaluation}.
In Section~\ref{section:related}, we review previous studies that have considered performance optimizations for deep learning, and finally, in Section~\ref{section:conclusion}, we conclude with a summary.

\section{Background}
\label{section:background}

\begin{figure}[t]
\center
\centering
\includegraphics[width=8cm, clip]{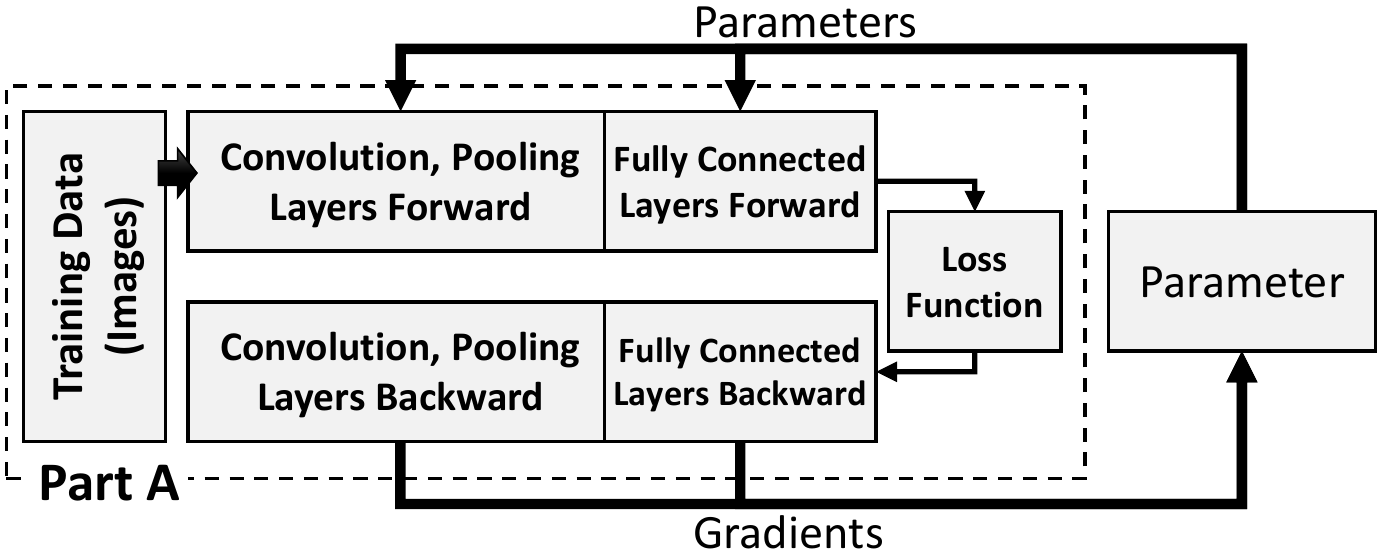}
\vspace{-0.2cm}
\caption{Structural convention of CNN}
\label{figure:traditional_arch}
\end{figure}

\Paragraph{Convolutional Neural Networks.}
Convolutional Neural Networks (CNN) is a class of neural network used in
a number of tasks ranging from image classification~\cite{2012alexnet, zhang2016augmenting, 2017inception4, 2015vgg, 2016resnet} to video recognition~\cite{xu2015discriminative, zha2015exploiting} and recommender systems~\cite{zhang2017deep}.
Figure~\ref{figure:traditional_arch} shows the workflow of conventional CNN model training and the various stages that it goes through during execution. 

With the training data (e.g., images) as input, model training proceeds through a forward pass and a backward pass in an iterative fashion.
First, the feature of the input data is extracted through
multiple layers in the forward pass of model training. 
Using the extracted feature, a loss function calculates the loss value, 
which is a similarity measure corresponding to the output value and classification.
Then, the backward pass, also referred to as back propagation, computes a model update
by going through the forward pass layers in reverse order.
During the backward pass, each layer obtains the
gradient, i.e., the change in the weights and biases (that is, the
parameters), and minimizes the loss value.
This process is repeated until the convergence of training loss.

Figure~\ref{figure:traditional_arch} shows that the forward-backward computation in CNN model training largely consists of convolution layers, pooling layers, and fully connected layers. 
The convolution layer processes the convolution
operation that extracts high-level features of the input data.
The pooling layer is an operation that reduces space in the horizontal and
vertical directions. It is common to place a pooling layer after the
convolution layer. Typically, there are two types of pooling, namely, max and
average pooling. Max pooling takes the maximum value in the target area
of the input data, while average pooling takes their average
value.
The fully connected
layer connects neurons of adjacent layers and can arbitrarily set the number
of outputs. 
As all the neurons are connected, they contain a large number of parameters. 
The last layer of a CNN model is generally a fully connected layer that outputs the predicted value.

\Paragraph{Distributed Deep Learning Training.}  
As deep learning models become more sophisticated and are trained on larger datasets, 
it is becoming common to scale training across machines in a distributed setting.
The most common form of such Distributed Deep Learning (DDL) training is \emph{data parallelism}~\cite{nsdi_tiresias}
where multiple workers train on their own set of data in parallel.
Essentially, each worker loads a complete copy of the model
(Part A in Figure~\ref{figure:traditional_arch}) into its own memory of an accelerator such as a GPU that is assigned to the worker.
In each training iteration (or step), each worker performs training using a subset of the input data
that is equally divided among all workers.
At the end of the iteration the workers exchange gradients to synchronize (or \emph{aggregate}) model updates.



A typical form of model aggregation is the parameter server (PS) architecture~\cite{li2014scaling}, which is popular in production systems~\cite{nsdi_tiresias, jeon_philly_2018}.
In this architecture, the PS hosts the master copy of the DDL model and is 
in charge of updating the model using the local results sent from all workers. 
The workers pull back the updated model from the PS at the beginning of each iteration and proceeds through the next iteration.
Note that the architecture itself does not limit the number of PSes that can be deployed~\cite{li2014scaling, peng2018optimus}.

\begin{figure*} 
\centering
\subfigure{
\includegraphics[width=0.235 \linewidth]{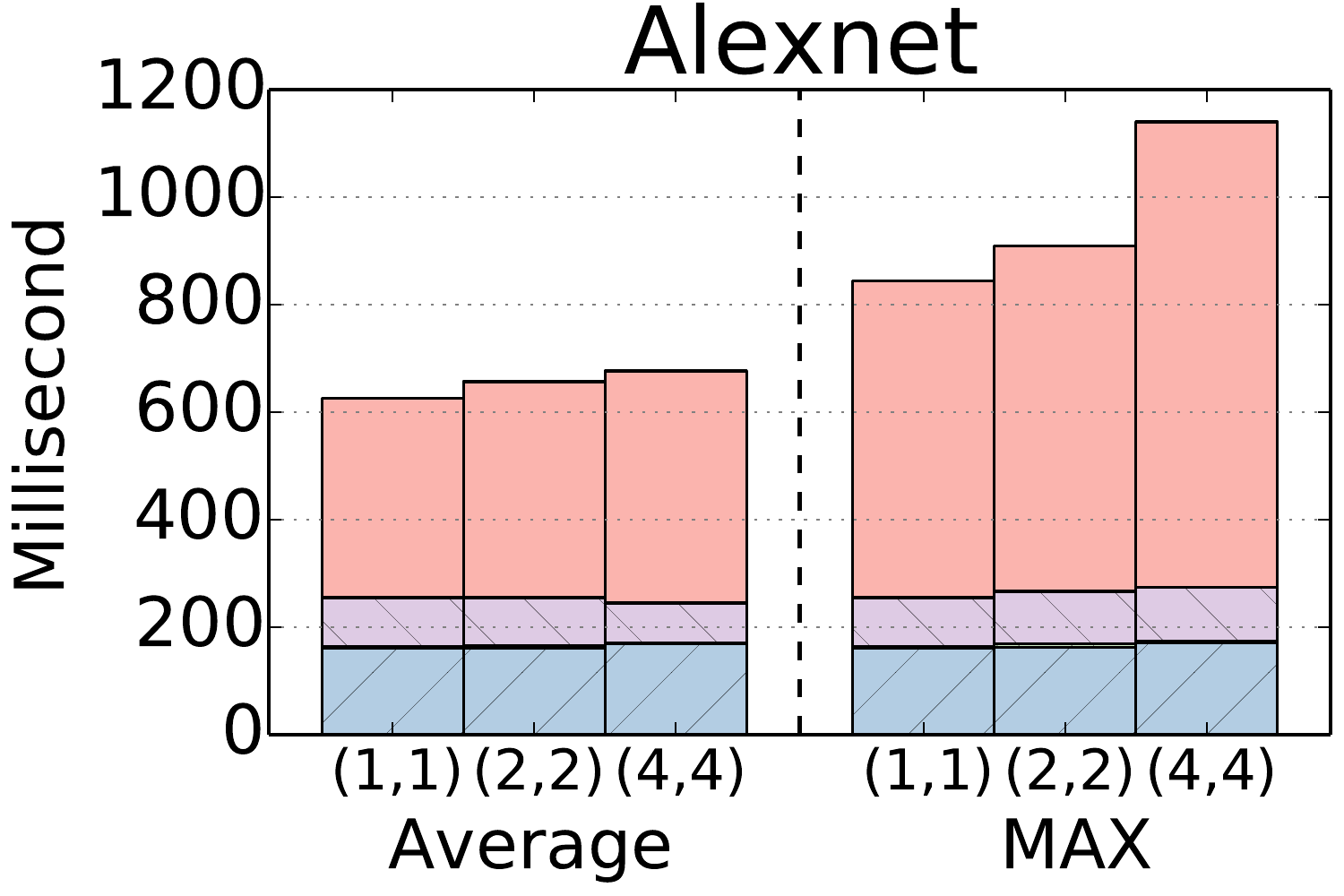}
}
\centering
\subfigure{
\includegraphics[width=0.235 \linewidth]{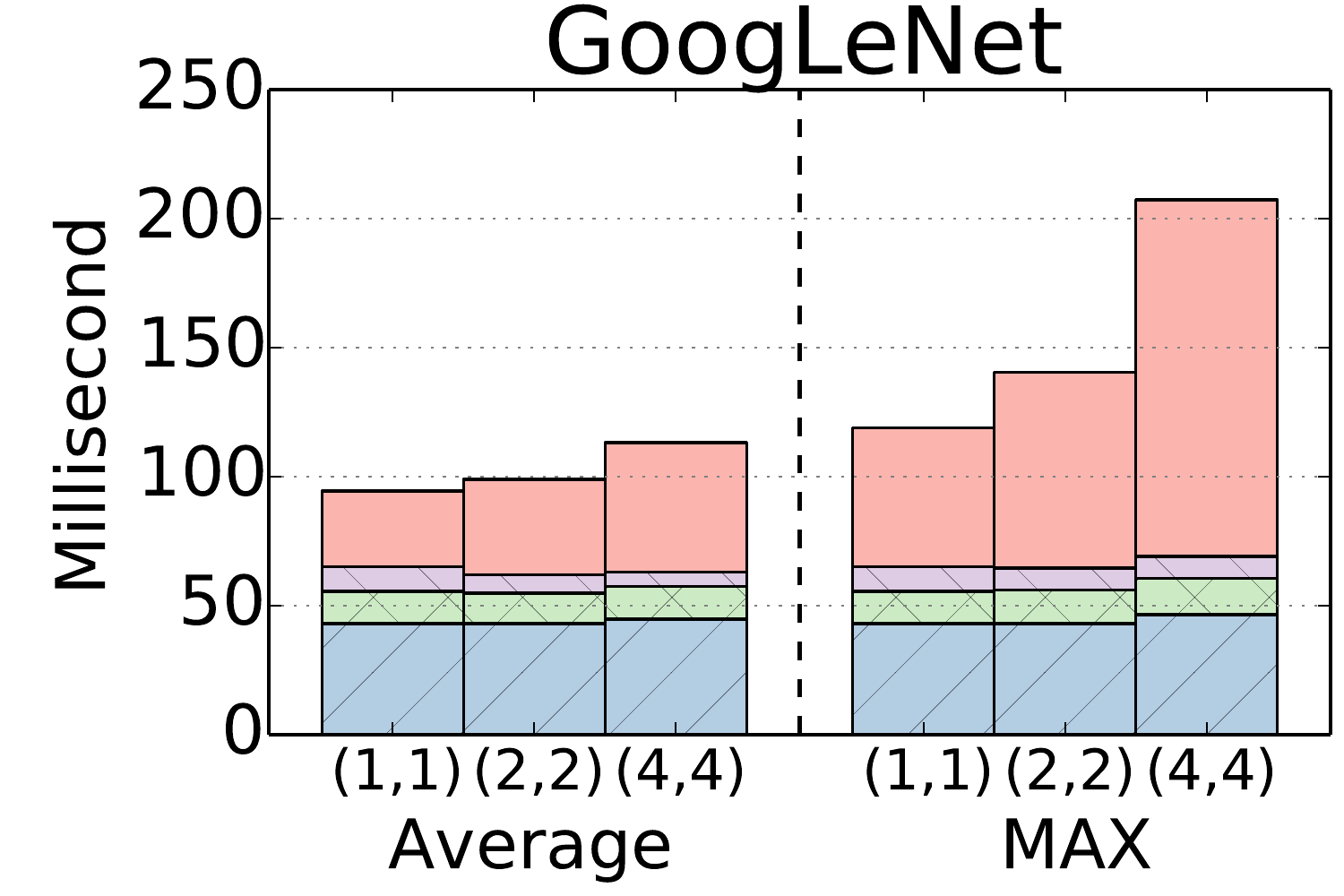}
}
\centering
\subfigure{
\includegraphics[width=0.235 \linewidth]{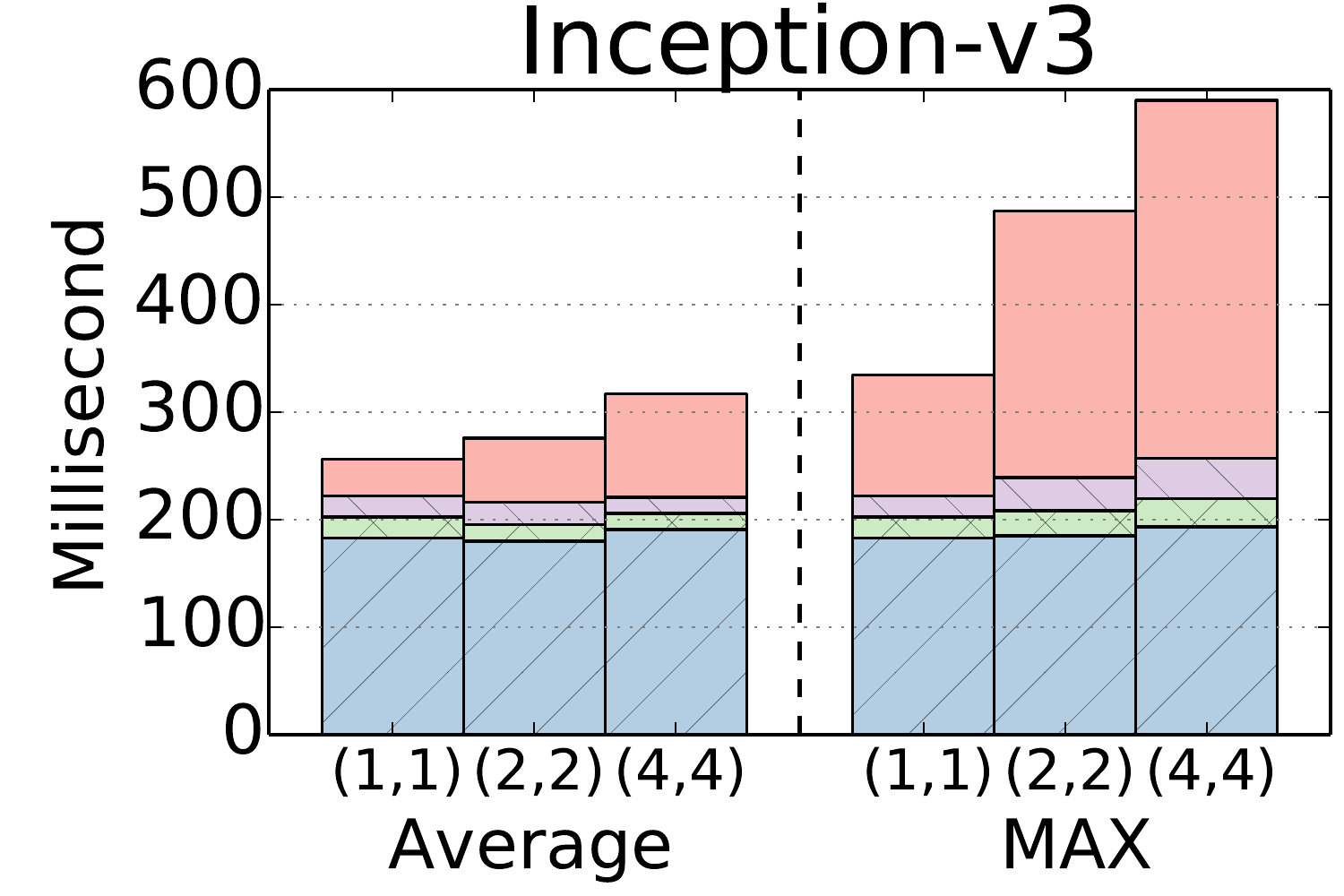}
}
\centering
\subfigure{
\includegraphics[width=0.235 \linewidth]{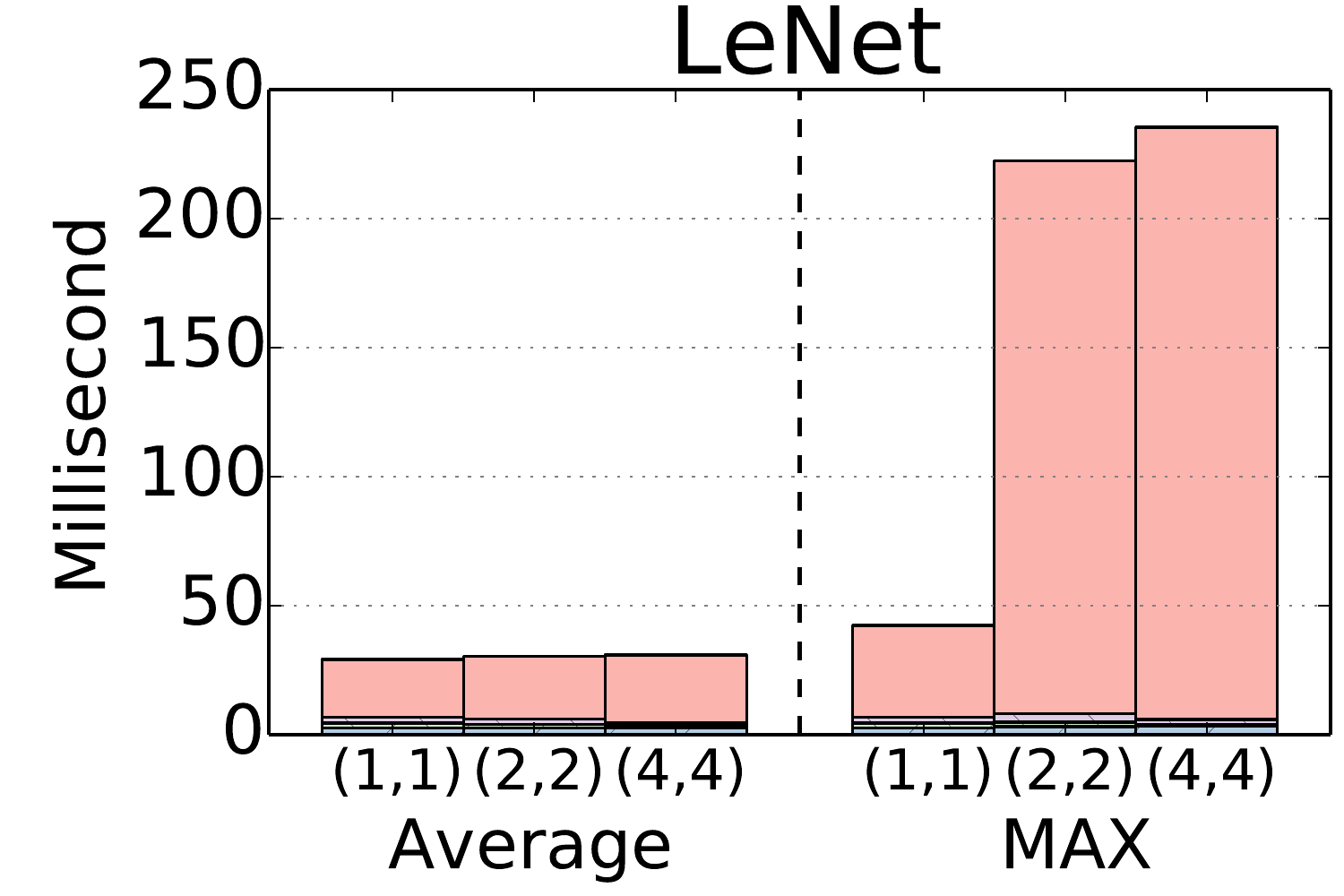}
}
\centering
\subfigure{
\includegraphics[width=0.235 \linewidth]{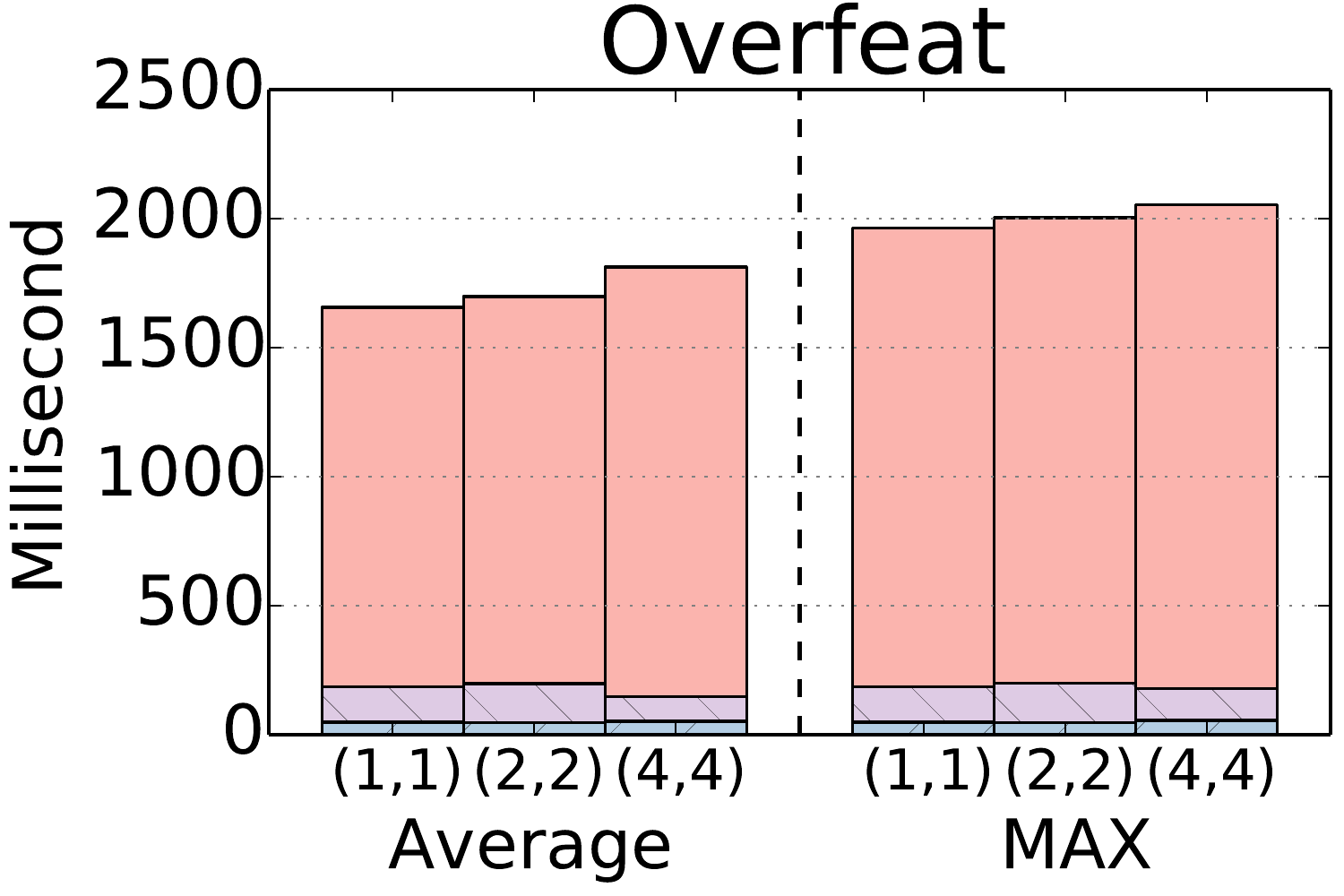}
}
\centering
\subfigure{
\includegraphics[width=0.235 \linewidth]{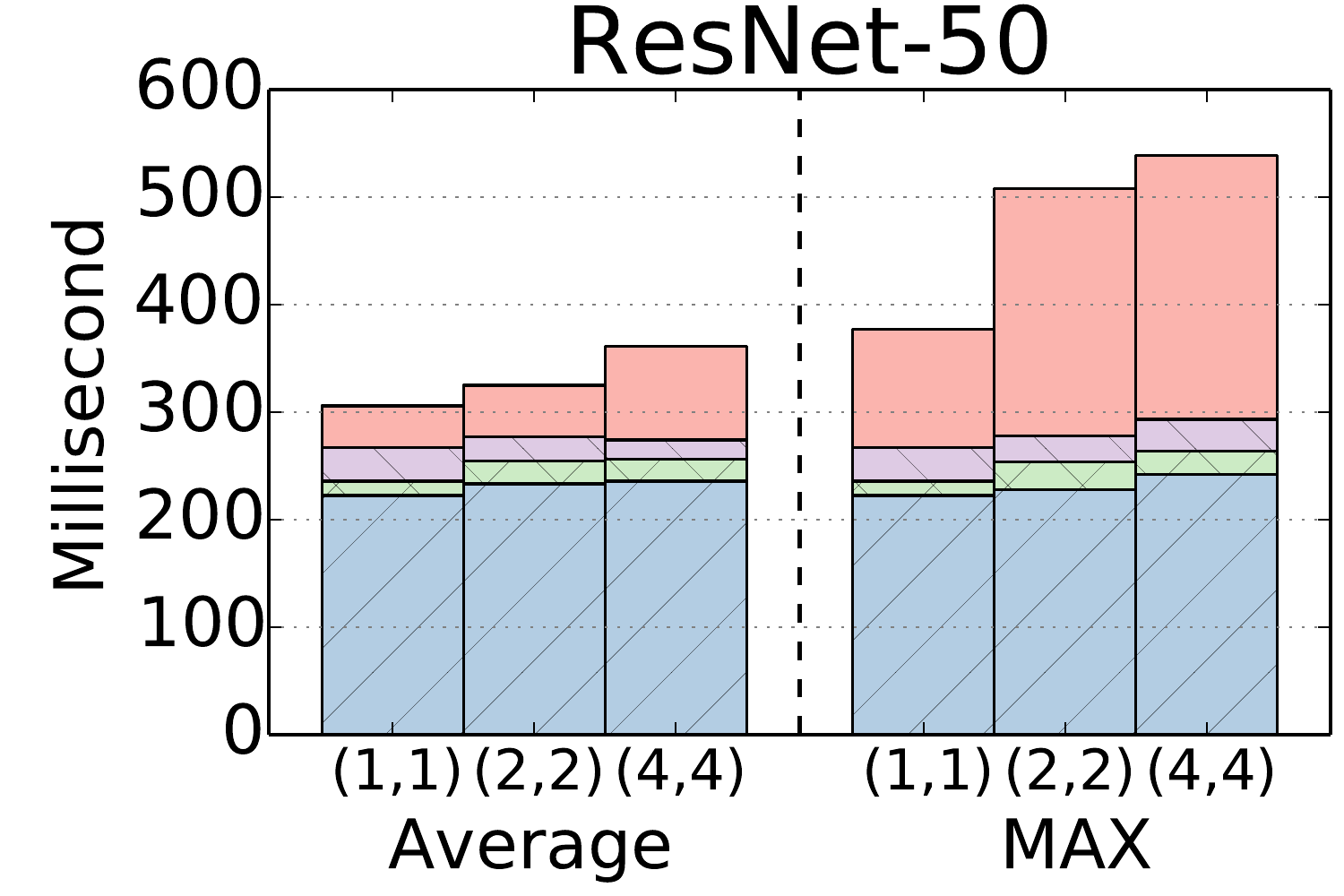}
}
\centering
\subfigure{
\includegraphics[width=0.235 \linewidth]{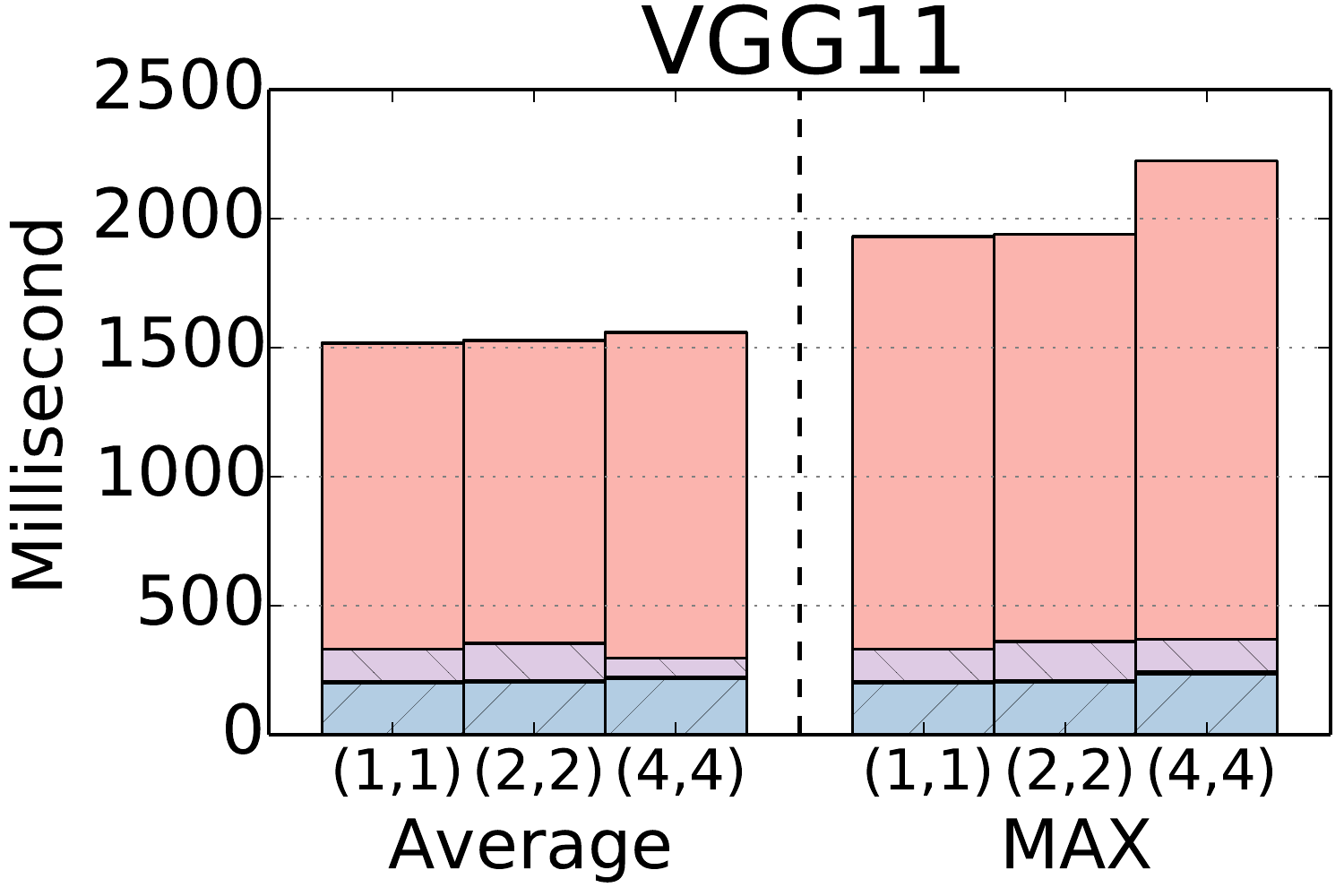}
}
\centering
\subfigure{
\includegraphics[width=0.235 \linewidth]{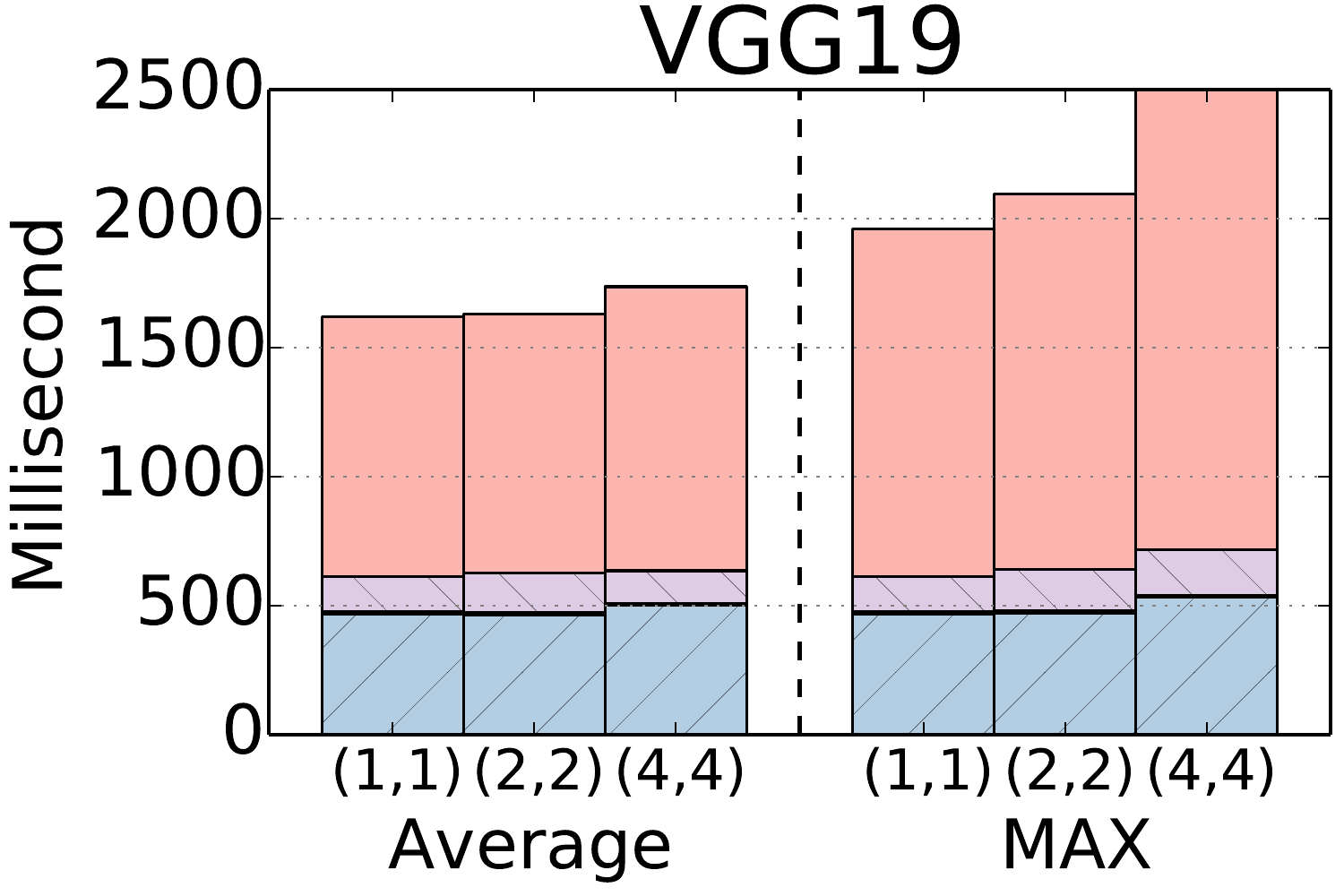}
}
\subfigure{
\includegraphics[width=13cm, clip]{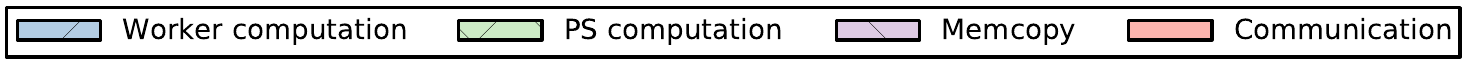}
}
\caption{Distributed training: time breakdown (after 100 step executions) including network transfer time, where the left bars are average step execution time for all workers, while right bars are the average step execution time for the slowest worker. Numbers in parentheses represent number of workers and PSes.} 
\label{figure:distributed_time_breakdown}
\end{figure*}

\section{Motivation: CNN Characterization}
\label{section:Characterization}

In this section we discuss the characteristics of training for CNN distributed deep learning (DDL) that stands as motivation behind the design of \sys{}, which we propose.
We focus our discussion on communication between workers and PSes during model aggregation and the computation involved at each worker.
For this motivation study, we use the CNN model benchmarks~\cite{tf_benchmakrs} that are supported in TensorFlow 1.12 and the ImageNet-1K dataset~\cite{imagenet_dataset}. 
The results reported in this section are based on experiments on TensorFlow 1.12 executed on a cluster of up to 8 machines each equipped with 4 NVIDIA TITAN Xp GPUs connected over a 56~Gbps RDMA network.
This is the same experimental setup used in Section~\ref{section:evaluation}, where we discuss it in detail.

\Paragraph{High communication costs in CNN DDL training.}
In CNN DDL training, the model aggregation phase is part of the critical path, having a significant hindering effect on training progress.
To illustrate, consider a job using data parallelism where all workers must communicate
with the PS to synchronize all the parameters of the model in each training step.
As the workers calculate the gradients, these values are sent to the PS.
Once the parameters are aggregated, these values are sent back to the individual workers.
That is, communication between workers and the PS occurs twice at every training step, and the amount of network transfer is proportional to the number of workers used in distributed training.

To assess this overhead, we perform a set of experiments, whose results are shown in Figure~\ref{figure:distributed_time_breakdown}, where we break down the end-to-end time spent for each worker to finish a training step, including the time spent for model aggregation. 
More specifically, the elapsed time is divided into four categories according
to the type of operation performed: (i) \textbf{worker computation} for model training by the worker;
(ii) \textbf{PS computation} for the aggregation step in the PS;
(iii) \textbf{memcopy} for copying data between the host machine and the GPU; and (iv) \textbf{communication} for transferring model parameters over the network.
The numbers reported for the first three operations are those provided by the TensorFlow timeline tracer, while the numbers of the last operation are obtained by measuring the end-to-end time for training and then, subtracting the sum of the first three values. 
We trained a total of 100 steps for each worker and obtained the average and max end-to-end times for each step.

While we report, on the left bars, the average of each step over all workers, as we train a total of 100 steps for each worker, we also present the average (over the 100 steps) times of the slowest worker, which is the one that essentially determines the performance when all workers need to synchronize training progress at each step~\cite{paleo_2017}.
Our measurements are based on changing the number of workers, denoted by the left number in the parenthesis, 
and we set the number of PSes (right number) to be the same as guided by prior work~\cite{peng2018optimus}.

Figure~\ref{figure:distributed_time_breakdown} shows that network
communication performed in model aggregation dominates for a number of scenarios across the benchmarks.
This happens
even when a single worker and PS is used, which ought to produce the least amount of network traffic.
In particular, from the breakdown, we observe that the average fraction of time used for communication is 53\% across the benchmarks.
As model training scales out, we see an increase in both the absolute amount of time required for training and the dominance of communication time.
For example, when using 4 workers and 4 PSes, communication makes up, on average, 58\% of the total training time per step, and it is worse for the slowest worker (in MAX) taking up as much as 92\% of the total time in LeNet (4,4).
Our observation here is that in order to deliver high performance distributed training, it is crucial to reduce the volume of network transfer.

\begin{figure*} 
\centering
\subfigure{
\includegraphics[width=0.235 \linewidth]{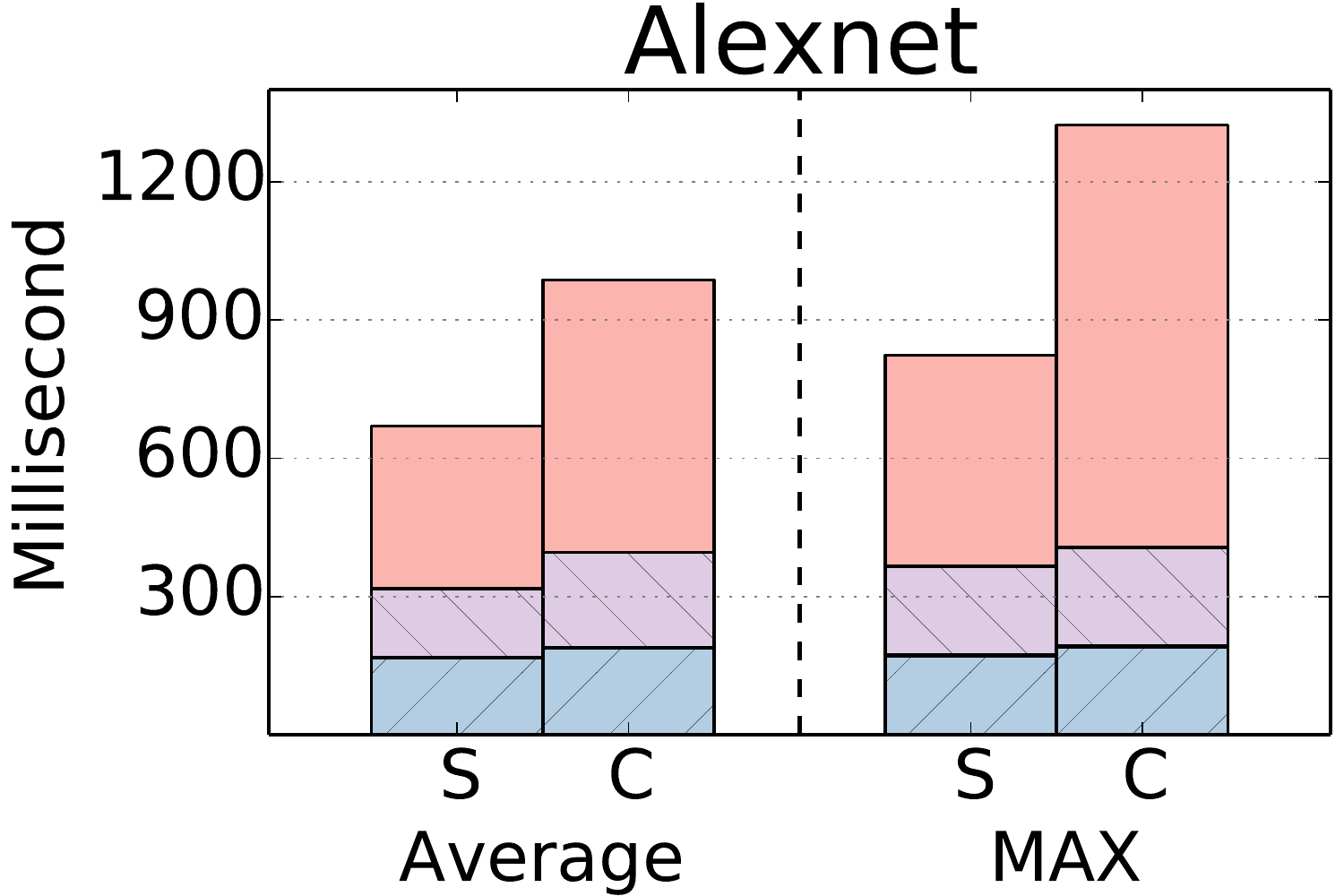}
}
\centering
\subfigure{
\includegraphics[width=0.235 \linewidth]{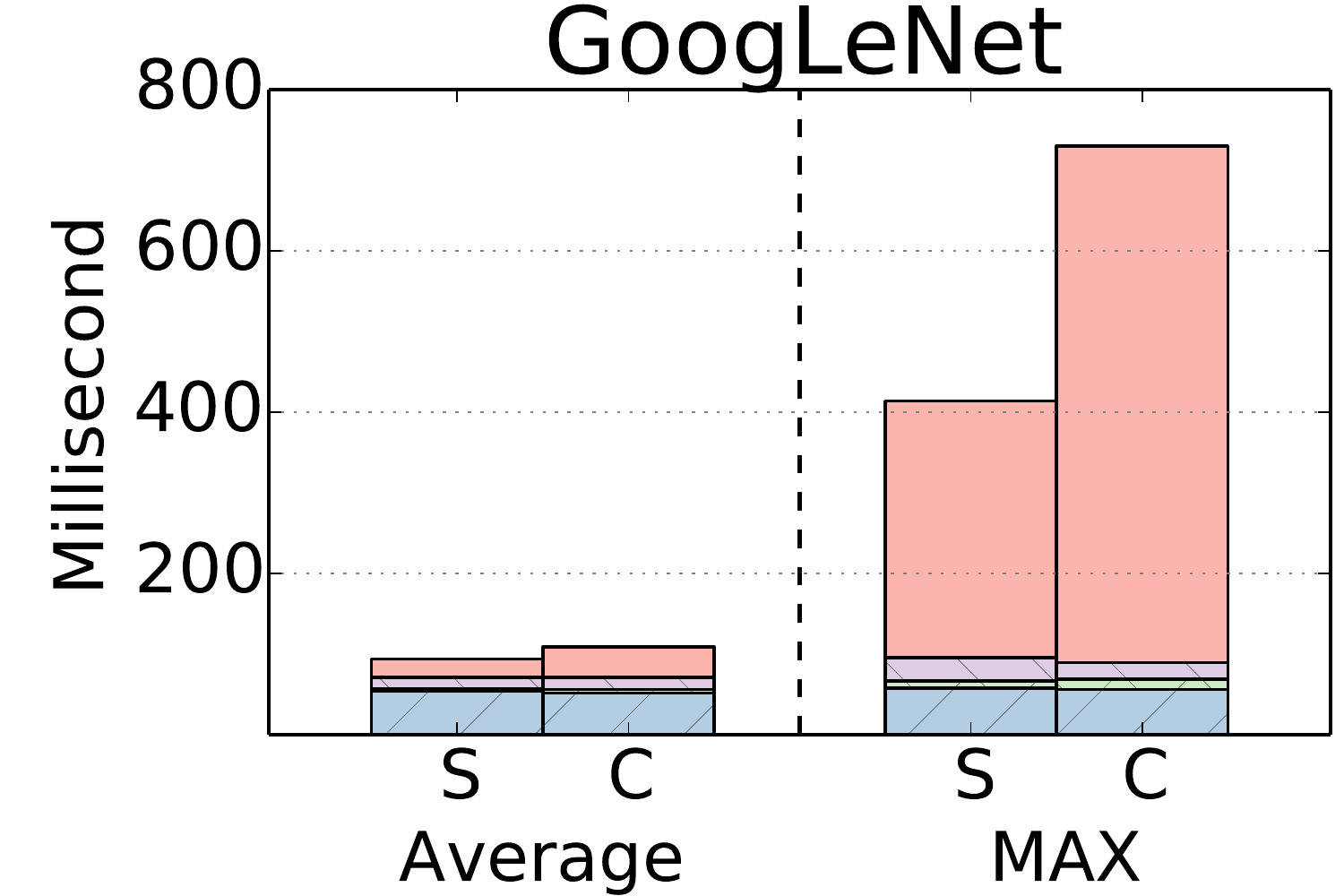}
}
\centering
\subfigure{
\includegraphics[width=0.235 \linewidth]{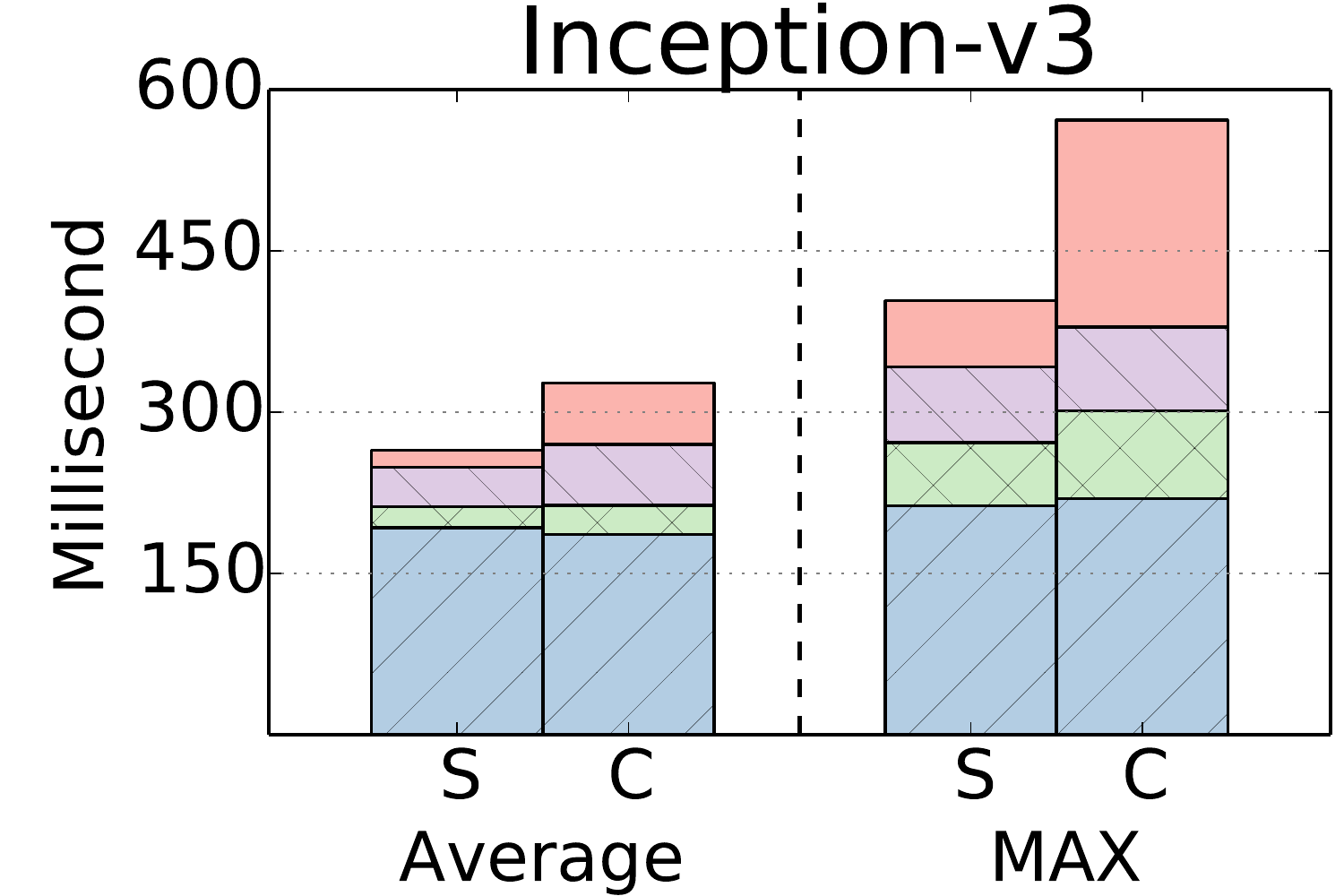}
}
\centering
\subfigure{
\includegraphics[width=0.235 \linewidth]{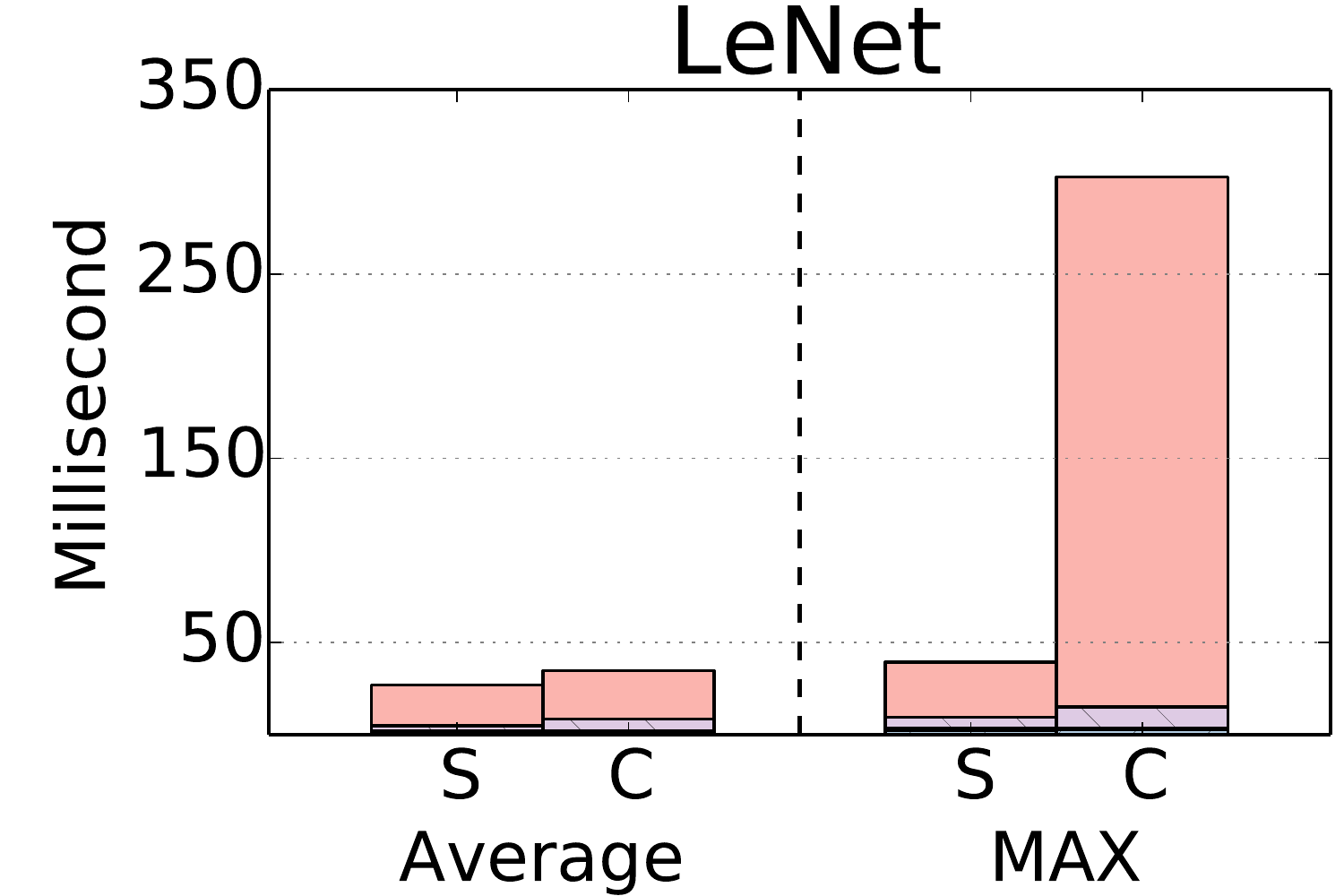}
}
\centering
\subfigure{
\includegraphics[width=0.235 \linewidth]{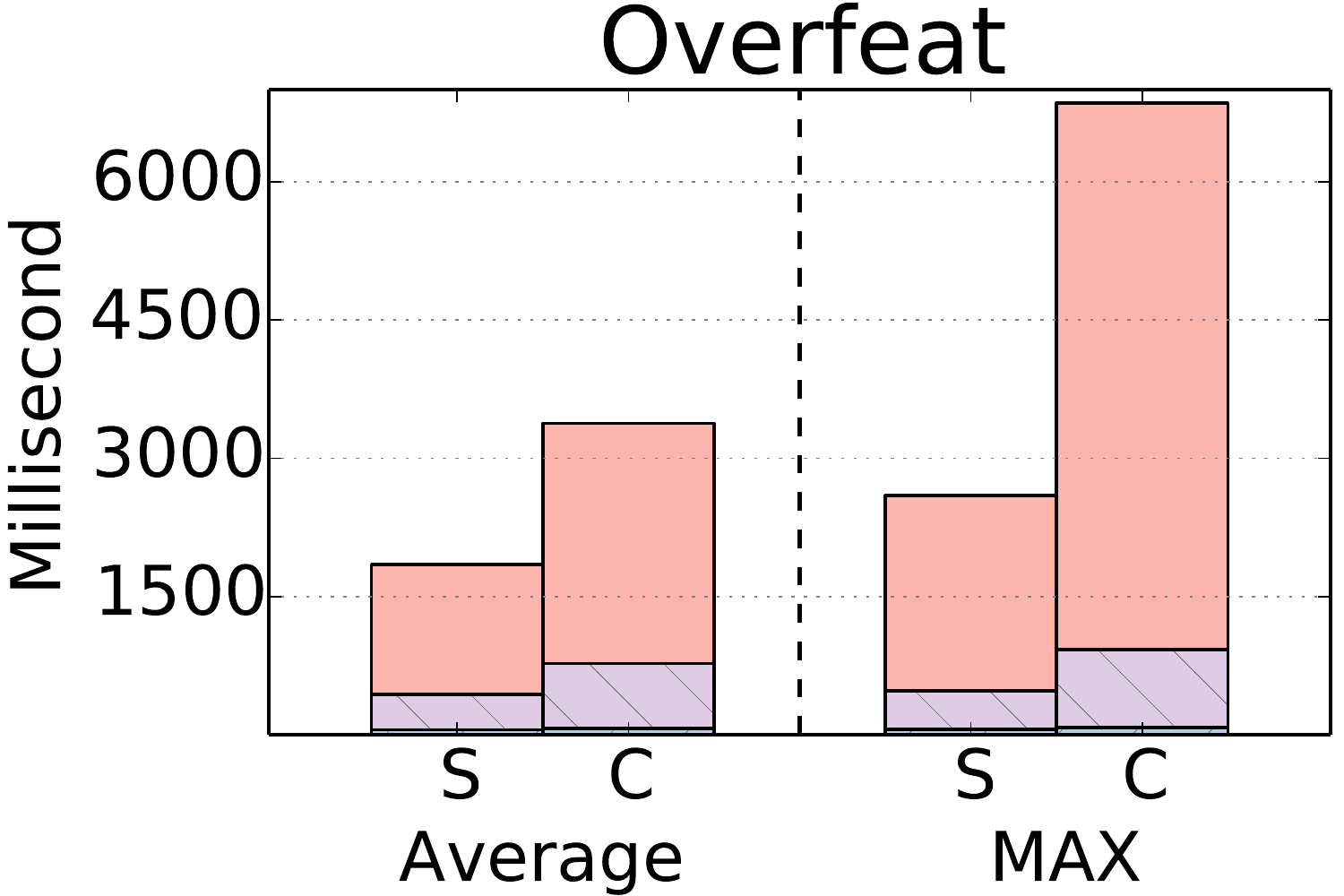}
}
\centering
\subfigure{
\includegraphics[width=0.235 \linewidth]{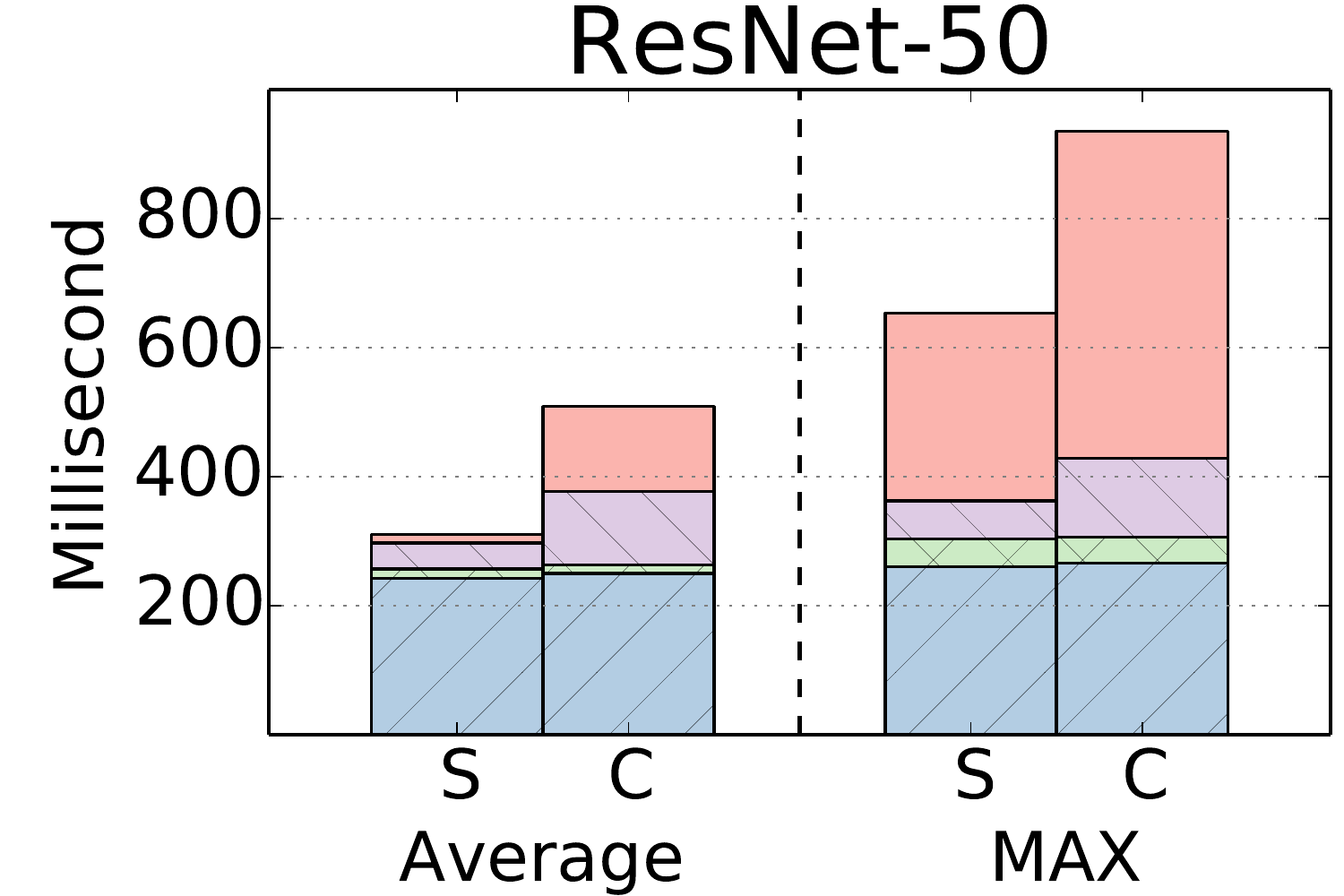}
}
\centering
\subfigure{
\includegraphics[width=0.235 \linewidth]{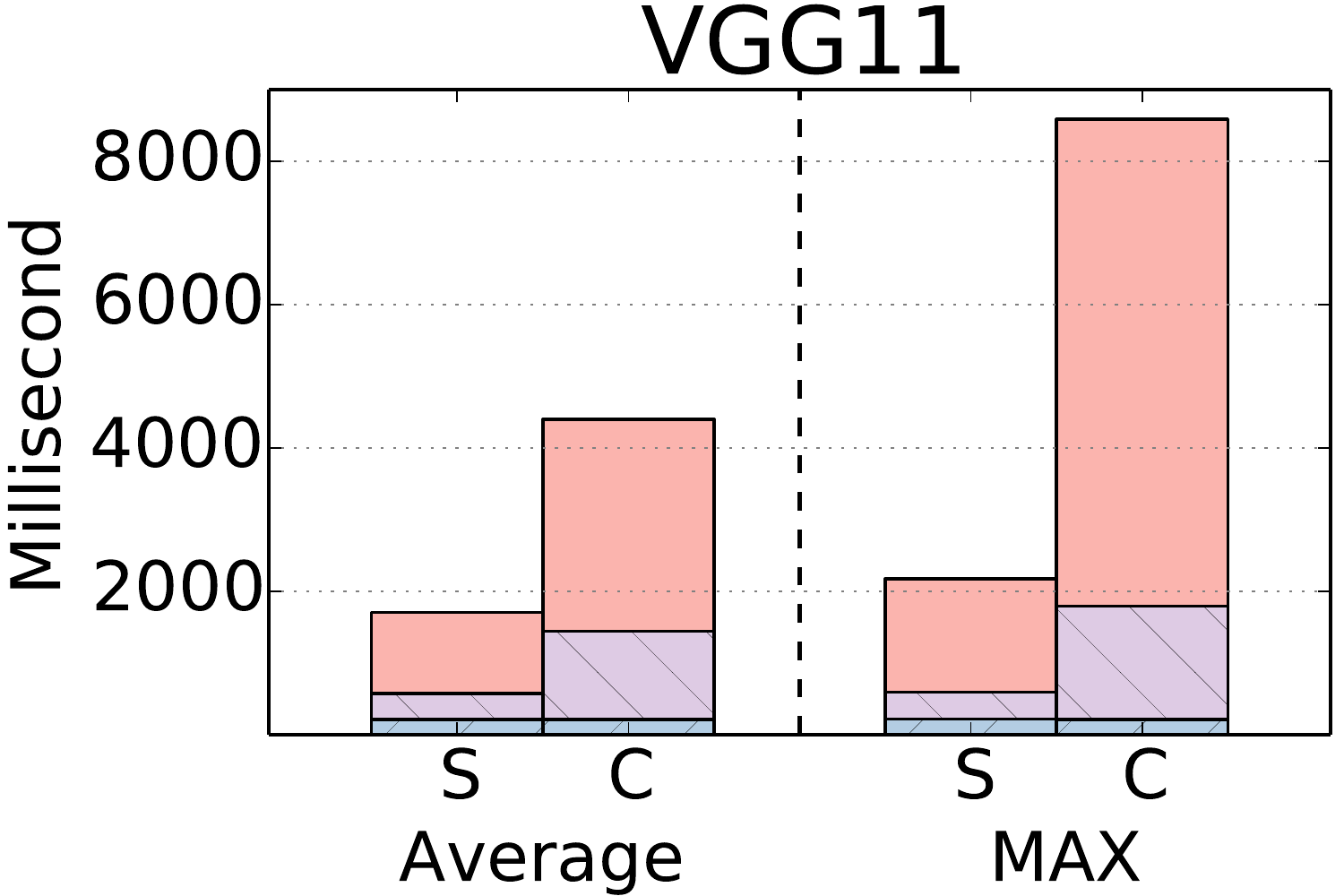}
}
\centering
\subfigure{
\includegraphics[width=0.235 \linewidth]{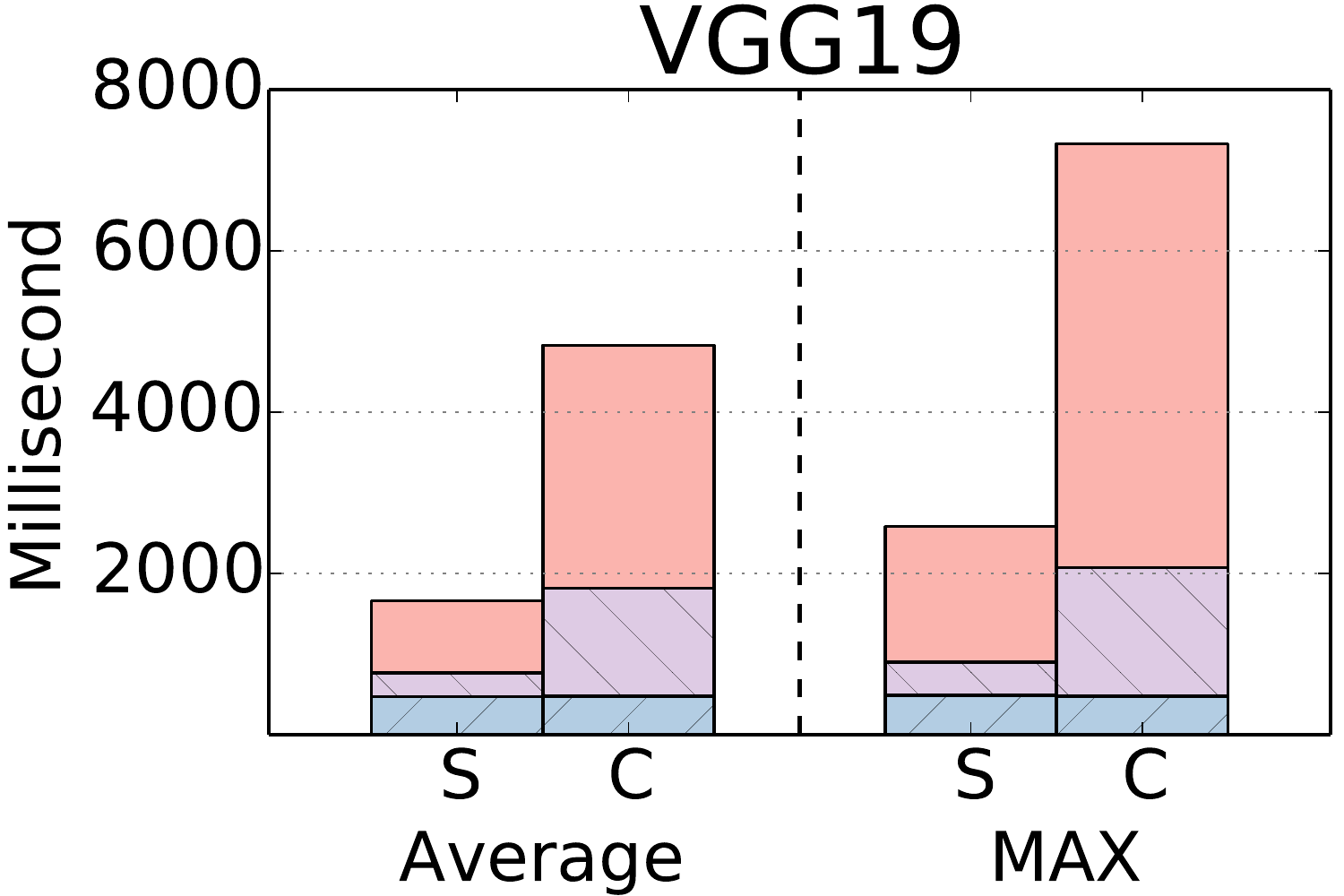}
}
\subfigure{
\includegraphics[width=13cm, clip]{fig/characterization/distributed_time/legend.pdf}
}
\caption{Network interference: S represents single model training, C represents consolidated training with 8 models in a cluster}
\label{figure:consolidation_time_breakdown}
\end{figure*}

\Paragraph{Network interference in shared servers.}
In a cluster of machines, consolidation of different jobs on the same cluster could lead to interference due to contention of shared network such as RDMA~\cite{xiao2018gandiva, nsdi_tiresias, jeon_philly_2018}. 
Such sharing could further deteriorate training performance that is already hampered by network communication overhead.
To confirm that workload consolidation indeed relates to further network interference and consequently, training progress, we make another set of measurements similar to Figure~\ref{figure:distributed_time_breakdown}, but this time on 3 worker, 1 PS training.
Figure~\ref{figure:consolidation_time_breakdown} compares the average execution time of 8 jobs that run concurrently on our 32-GPU cluster (denoted by C)
to that of a job that runs in isolation (denoted by S).

The results in Figure~\ref{figure:consolidation_time_breakdown} confirm that jobs further interfere with each other in the network, and that interference among jobs on shared resources has considerable effect, 
slowing down training time, in our experiments, by over $7\times$ for LeNet.
We also see that this inefficiency mostly comes from the time prolonged during model aggregation due to increased network contention. 
Thus, we conclude that optimizing model aggregation could improve performance not only when training is isolated, but also when
workload is consolidated to share resources.

\begin{figure}[t]
\center
\centering
\subfigure[Communication parameter]{
\includegraphics[width=3.9cm, clip]{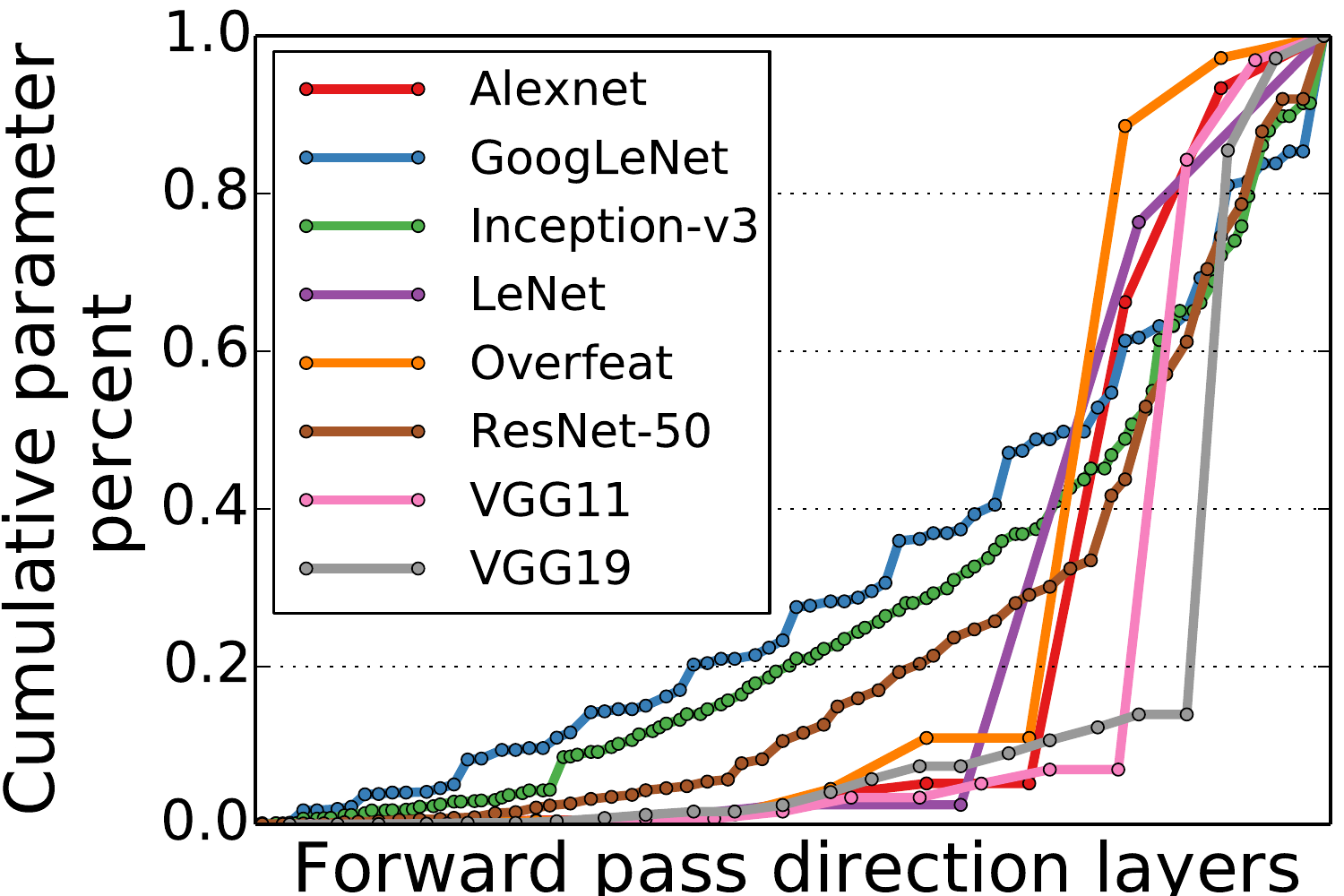}
}
\label{figure:parameter_cdf}
\centering
\subfigure[Computation time]{
\includegraphics[width=4.1cm, clip]{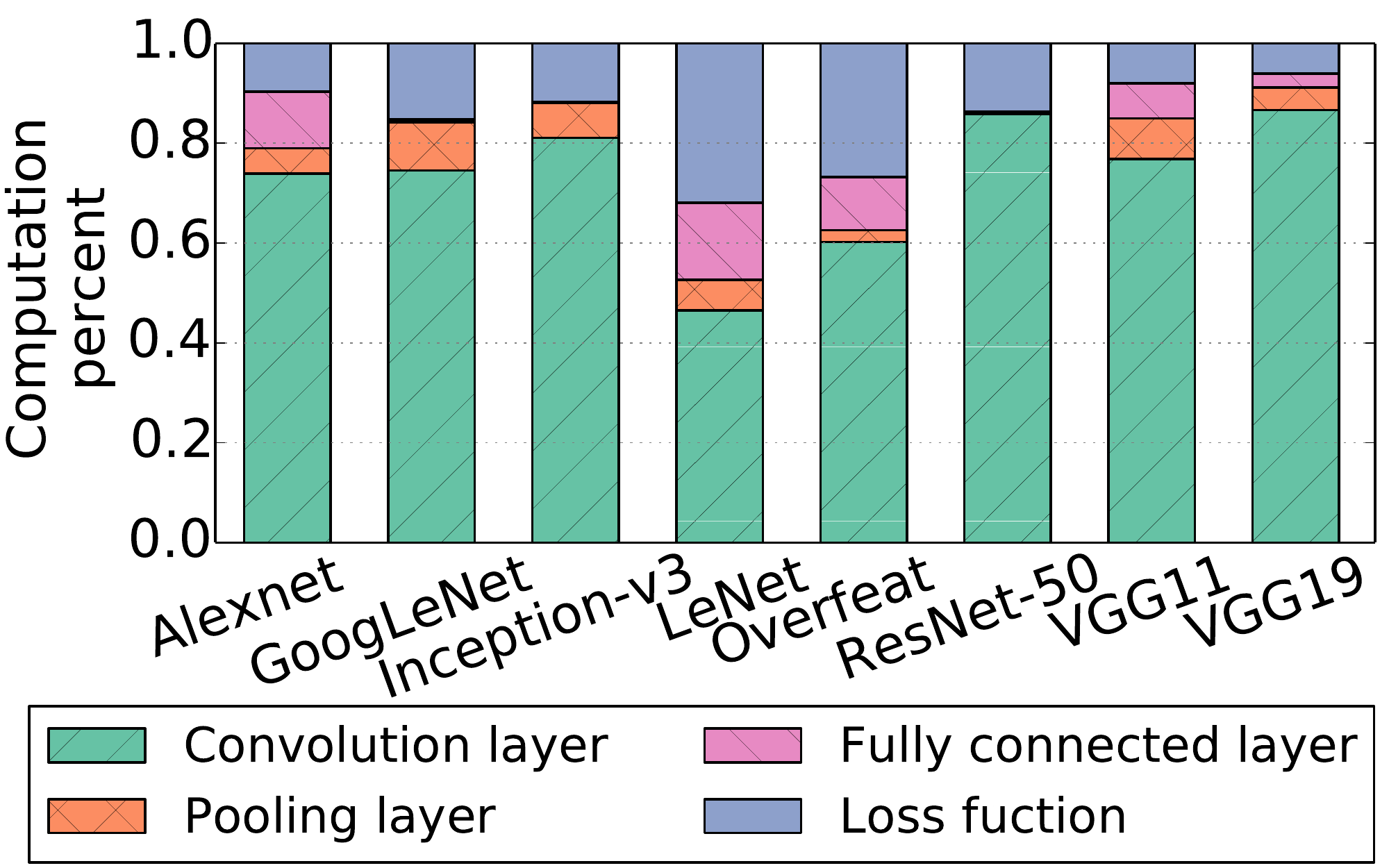}
}
\label{figure:computation_cdf}
\caption{Memory usage and computation of CNN}
\label{figure:conventionalCNN}
\end{figure}

\Paragraph{Disproportional Memory Usage and Computation.}
Another key characteristic of CNN
relates to how parameters and computation are laid out among the layers that comprise CNN. 
We now elaborate on these points.

First, for a number of CNN models, a significant fraction of memory is used by
a few layers exercised in the last phase of the training's forward pass.
Figure~\ref{figure:conventionalCNN}(a) shows the cumulative
distribution of memory usage in terms of parameter sizes in the order of the forward pass layers for each of the models.
Overall, the figure shows that a majority of CNN models have skewness in memory usage, with the last 25\% of the layers occupying more than 86\% of memory space in 5 out of 8 models.
On inspecting these memory-demand layers by layer type, we find that they correspond to
the fully connected layers.
Thus, there is an opportunity to reduce network transfer during model aggregation by placing
the fully connected layers within the PS machine.


Let us now consider the computation distribution in CNN.
Figure~\ref{figure:conventionalCNN}(b) shows the breakdown of computation time in CNN training by layer type. 
We see that a large fraction of the computation time is occupied by the convolution layers. 
Specifically, the convolution layers account for around 80\% of the total GPU time used during training on average. 
This characteristic makes colocating the fully connected layers with the PS a more promising approach,
because those computation-demand convolution layers mainly appear in the first phase of training's forward pass and do not
interfere with computation on the fully connected layers.
Note that it may be possible to accelerate the execution of the convolution layers by assigning more compute resources (e.g., GPUs) to these layers separately through layer partitioning.


\Paragraph{Challenges.}
In this section, we have shown that network communication is a major performance factor in CNN training whether the model is trained in isolation or in consolidation with others.
We also showed that a large proportion of the parameters, which need to be communicated to the PS, are concentrated in the latter layers.
This provides an opportunity to reduce network overhead by placing these layers within the PS's machine.
Also, we showed that the convolution layers are concentrated in the front phase of training, and thus, again, there is an opportunity to optimize computation.

However, blindly offloading the latter layers to the PS machine may not pay off for all CNN models as some of these models do not entirely correspond to the structure shown in Figure~\ref{figure:traditional_arch}.
For example, for some CNN models, the fully-convolutional network replaces the fully-connected layers with convolutions to take in an arbitrary input size for efficient inference and learning~\cite{long2015fully}.
Also, in the ResNet model, the network has many filters in the convolutional layer, and hence, many parameters, which requires sizable memory space~\cite{2016resnet}.
In these cases, either there is no significant skewness in parameter distributions among layers
or the earlier part in the CNN pipeline could consume
considerable memory, thus affecting the synchronization cost.
The main goal of our work is to provide a systematic way to identify models that could benefit from \sys{},
and structure a pipeline of layers across distributed resources 
to obtain higher training performance.


\section{Layer Placement with \sys{}}
\label{section:RALP}




We propose \sys{} --- Resource-Aware Layer Placement --- a scheme whose target is to reduce network communication and balance computation to expedite distributed training of CNN.
To achieve the goal, \sys{} carefully exploits workload characteristics
studied in Section~\ref{section:Characterization}. \sys{} places CNN layers,
which is composed of a communication and computation sequence, in particular machines in a way that data flows efficiently and compute-bound layers are executed in parallel.
Typically, a deep learning model is composed of a number of layers, and
it is feasible to place each individual layer anywhere among compute resources such as GPUs. 
However, since our focus in this paper is on CNN, we actively utilize
CNN-specific properties related to the model structure (shown in Figure~\ref{figure:traditional_arch})
in the optimization.
Nonetheless, we will briefly discuss how to apply \sys{} beyond CNN at the end of this section.


Since network communication is on the critical path of training in distributed mode and memory-demand layers are the root cause, \sys{} takes an approach that colocates those layers with the parameter server (PS) in the same machine.
In this way, the cost associated with model aggregation can be considerably reduced.
Further, \sys{} decides for a given model if the model could
benefit sufficiently from our resource-aware placement, and if so which "specific" layers
it needs to offload to the PS machine to have most benefit.

When deciding layer placement, \sys{} also factors in 
such layers that are computationally expensive.
To illustrate, consider a naive approach that places all layers with the PS.
Evidently, this is not a good solution as this will incur considerable
computational overhead on the PS machine. The natural choice for layer placement would be
to place the layers with small computation, but with large memory demands (i.e.,
parameters) in the
PS machine. This will have the effect of reducing network communication
without overloading the machine that runs both PS and the offloaded layers. 

\begin{figure}[t]
\center
\centering
\includegraphics[width=8cm, clip]{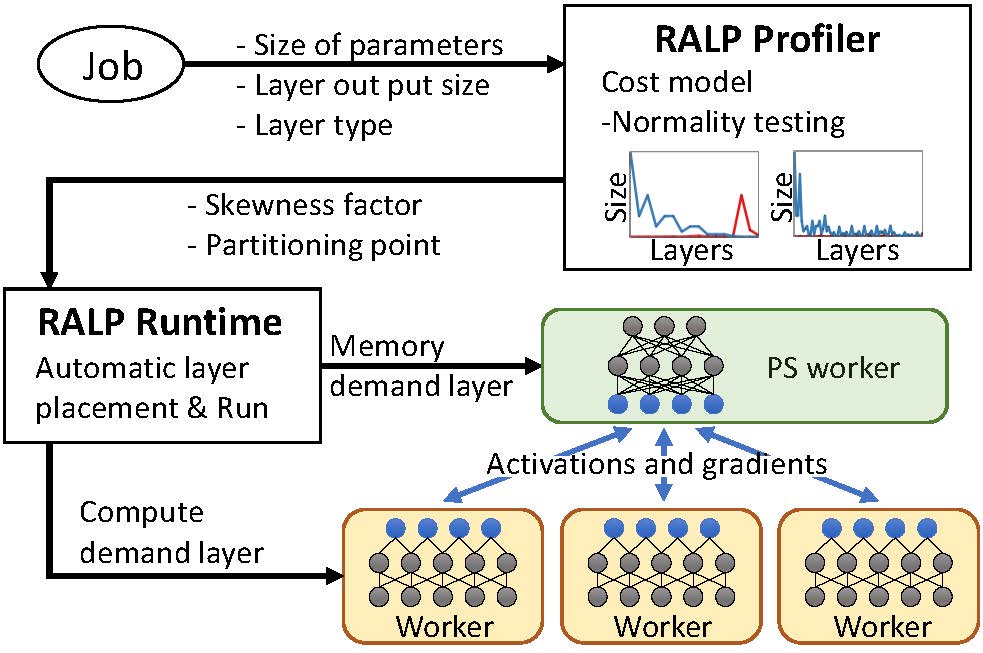}
\caption{Workflow of resource-aware layer placement for distributed training in \sys{}}
\label{figure:ralp_architecture}
\end{figure}

\subsection{System Overview}

Figure~\ref{figure:ralp_architecture} depicts the overall architecture of \sys{} and its two key components, RALP Runtime and RALP Profiler,
that work in complement with each other within the existing machine learning frameworks such as TensorFlow.

(i) \textbf{\sys{} Runtime} generates a distributed training plan and executes it on
distributed compute resources.
Training in \sys{} differs in several significant ways from
the traditional PS-based approach. 
Previously, as the number of workers increases, it is common practice to use the same number of PSes~\cite{peng2018optimus,nsdi_tiresias}.
This is partly because network transfer is the bottleneck and
using few PSes could make the PSes overwhelmed by the large volume of data to receive, process, and send back.
In \sys{}, we can
initiate distributed training efficiently using a smaller number of PSes for two reasons:
(i) network transfer is no longer the bottleneck, and (ii) the
PS machine that includes the memory-intensive layers (e.g., the fully connected layers) 
is not compute heavy~\cite{2014OWT,2018JiaExploring}.
In effect, \sys{} can save resource usage to train a job while using the same number of workers,
which mostly run the compute-intensive layers (e.g., the convolution layers).

(ii) \textbf{\sys{} Profiler} profiles the input CNN to estimate network communication across layers, and then informs where to \emph{partition} between the layers to be network-cost effective.
Blindly applying \sys{} to every CNN model will deteriorate some models that do not result in a sizable reduction in data transfer.
Thus, our profiler uses a cost model that informs if there exists such gain through partitioning.
Essentially, the cost model exploits the parameter size for each layer and additionally the size of activations and gradients exchanged between two neighboring layers during the forward-backward pass.
In addition, the profiler takes computation costs into account in a simple yet effective way based on CNN's characteristics: we avoid partitioning layers when the optimal point for splitting the model is found 
between the convolution layers, which are computationally costly.
This makes the profiler widely applicable as it depends only on the model itself including training configuration
(e.g., batch size), not on dynamic factors such as clock speed of in-use GPUs.

\Paragraph{Workflow.}
Figure~\ref{figure:ralp_architecture} shows the stages that \sys{} goes through.
Once a job is received by \sys{}, it is delivered to \sys{} Profiler,
which determines if partitioning is adequate or not. 
Jobs can bypass the profiler if the characteristics of the jobs are already known, or has been estimated offline, to benefit from partitioning.
The profiler consults the cost model to determine the eligibility of applying \sys{}.
Only if a job is predicted eligible, the model is partitioned and placed
across particular machines; otherwise, no partitioning occurs, and the model training runs on the standard
PS architecture.

In the subsequent sections, we explain distributed training in \sys{} Runtime and \sys{} Profiler in more detail.


\subsection{Distributed Training}

Figure~\ref{figure:ralp_architecture} also shows the training procedure and communication of distributed training enabled by \sys{} Runtime. 
For simplicity, our explanation is based on one PS and multiple workers, a default configuration on a moderate
degree of parallelism (e.g., 32 workers).
We further discuss how to enable multiple PSes in \sys{} in case training must scale on much higher degree.
We henceforth use the term \emph{PS worker} to explicitly denote the training procedure ongoing in the PS machine.

\Paragraph{Training procedure.}
Workers in \sys{} run computation-demand layers and require enough compute resources to speed up the training progress.
We follow conventional data parallelism to run workers concurrently.
Workers must communicate with the PS worker independently during ongoing forward-pass and backward-pass computations.

In each training step, the PS worker receives activations from all workers, which arrive at the PS worker in an arbitrary order. 
In order to make sure that the backward pass computes the gradients using the same version of weights accessed
in the forward pass, the processing of incoming activations is not interleaved, i.e., one batch of activations
from a worker at a time.
As the memory-demand fully-connected layers in the PS machine are
not computation-intensive~\cite{2018JiaExploring}, 
this processing is typically fast and does not add back pressure.

\Paragraph{Model aggregation.}
We already explained how workers communicate with PS workers during training,
and we now describe model aggregation occurring at the end of each training step.
The most notable distinction from the original PS architecture is that model aggregation for the memory-demand layers is done through intra-machine data transfer.
Workers may need to synchronize with the PS over the network,
and thus, in this case, the amount of parameters to transfer over the network is small.
This synchronization is again fast, alleviating pressure on network usage.

\Paragraph{Enabling massive parallelism.}
In our current evaluations, we only consider a single PS worker configuration, where we observe no performance issue while fully exploiting all other GPUs in our 32 GPU testbed to run a single training job.
Even so, given that the largest job reported in a production cluster runs on 128 GPUs~\cite{nsdi_tiresias}, we plan to support multiple PS workers to balance the incoming load.
To facilitate multiple PS workers, we require that multiple PS workers run in parallel while aggregating activations, and that those PS workers partition computation previously done in the single PS worker.
This is an interesting direction for further investigation, and we leave this part of the study as future work.

\Paragraph{Synergy with resource manager.}
For distributed training, most deep learning frameworks require all GPUs be available at the same time~\cite{nsdi_tiresias,jeon_philly_2018}.
In traditional resource scheduling in a shared cluster, one of the biggest concerns has been on resource placement such that locality becomes as high as possible, i.e., packing the job's GPUs within as fewer machines as possible~\cite{nsdi_tiresias,jeon_philly_2018}.
This is because greater locality improves training time by bringing down model synchronization overhead.
However, in practice, such locality constraints often need to be relaxed to reduce waiting times, especially for jobs that use many GPUs~\cite{jeon_philly_2018}.
With RALP, the scheduler can facilitate relaxed locality without sacrificing much training performance as a training job can eliminate a significant fraction of network transfer.

\subsection{Model Profiler}
\label{section:RALP_profiler}

We now present a model profiler that decides resource-aware layer placement at runtime on 
each training job.

\begin{figure}[t]
\center
\centering
\subfigure{
\includegraphics[width=4cm, clip]{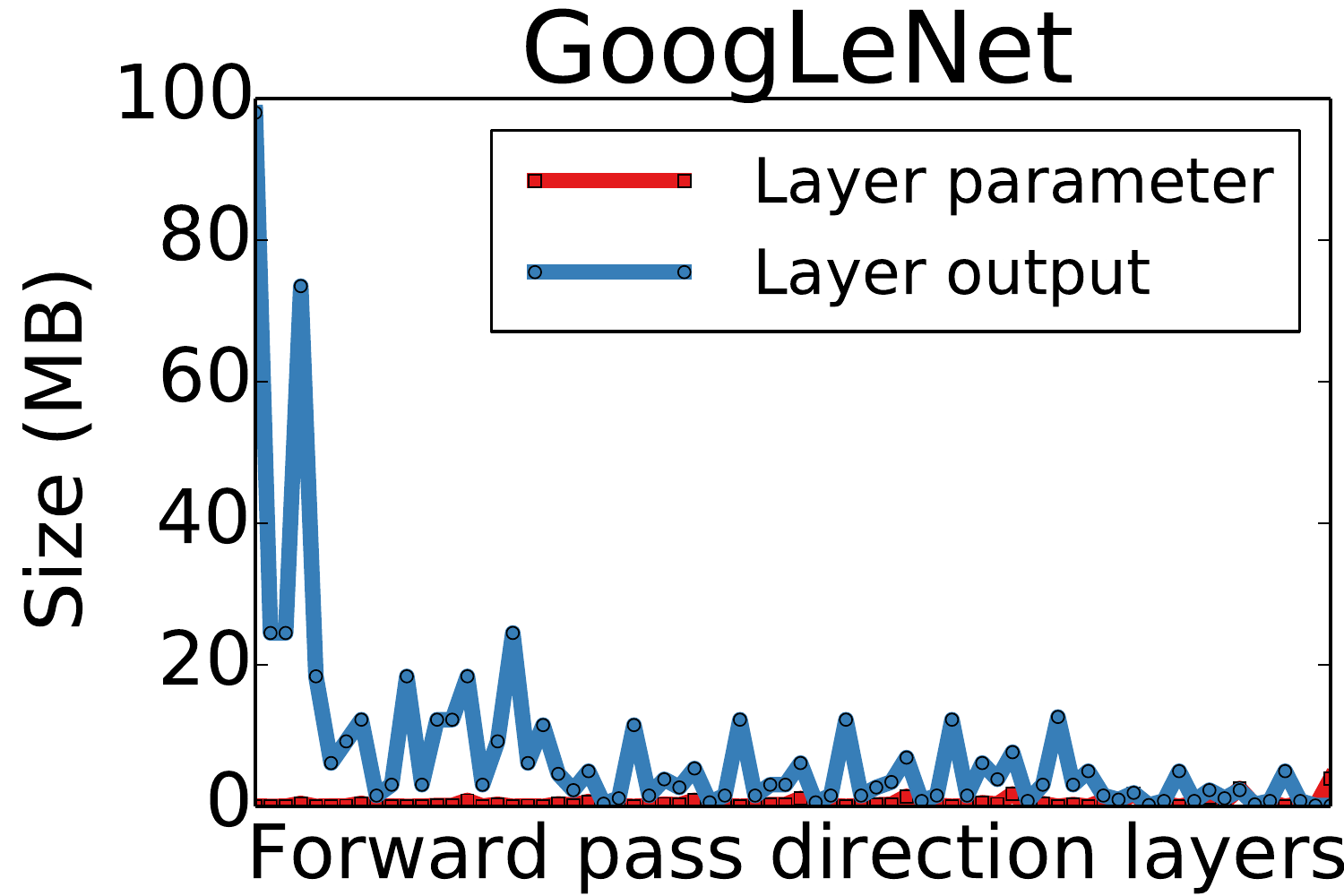}
}
\centering
\subfigure{
\includegraphics[width=4cm, clip]{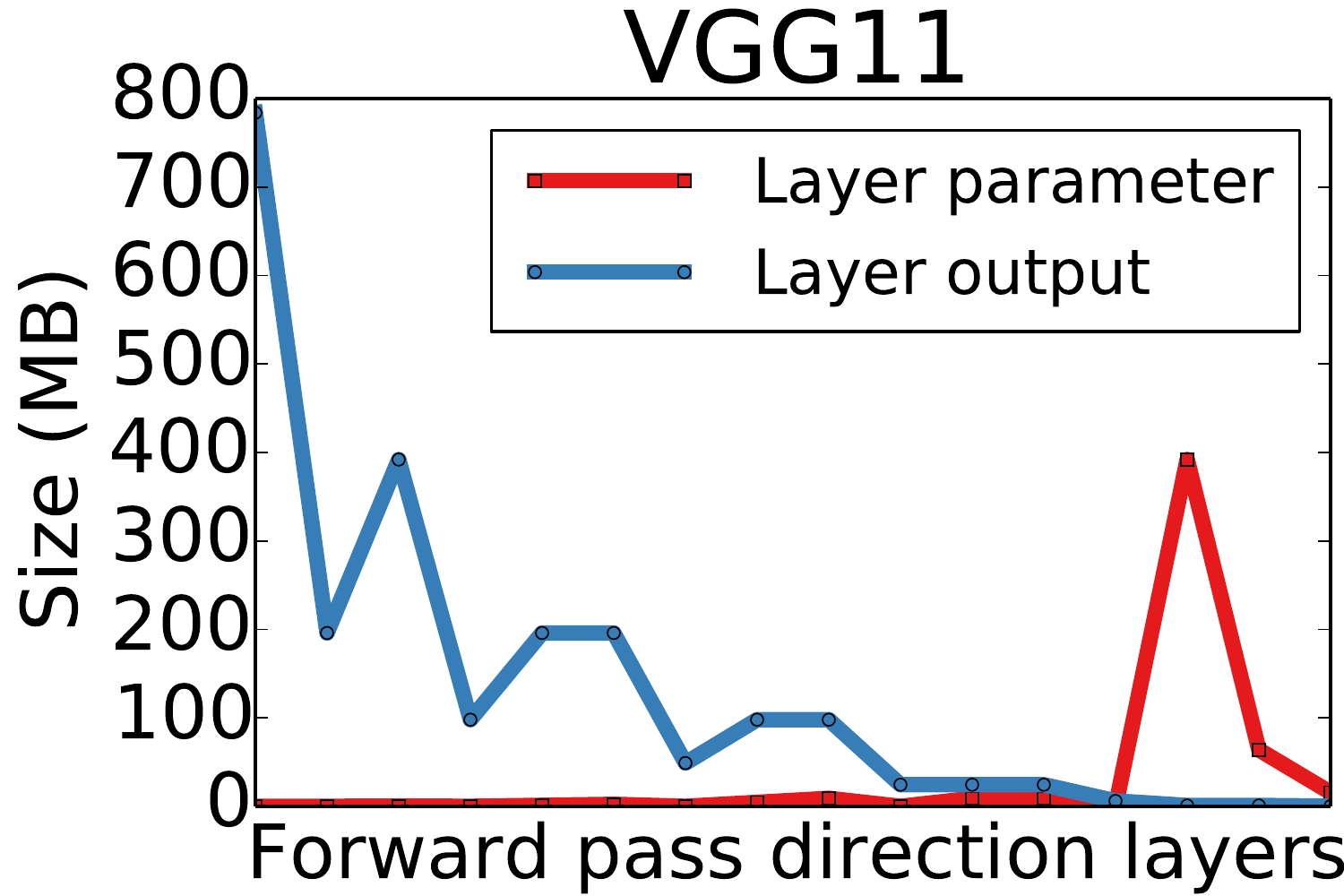}
}
\caption{Distribution of layer and output sizes across layers}
\label{figure:skewness}
\end{figure}

\Paragraph{Inputs.}
Our model profiler takes three inputs to decide on the layer placement:
(i) the size of parameters used in each layer, 
(ii) the network traffic each layer would
invoke, that is, the output size of the layer,
and (iii) the type of each layer.
The output size of the layer is proportional to the batch size,
which is the number of images used in one step of training.
Note that in CNN, the output size tends to decrease in the order of
the forward pass layers as shown in Figure~\ref{figure:skewness}.

\begin{table*}[]
\centering
\caption{Parameter skewness factors for 8 benchmarks under our study: larger absolute values indicates higher skewness.}
\begin{tabular}{l||r|r|r|r|r|r|r|r}
\hline
                  & Alexnet & GoogLeNet & Inception-v3 & LeNet & Overfeat & ResNet-50 & VGG11 & VGG19 \\ \hline \hline
Skewness factor & $-2.27$   & $-0.74$     & $-0.96$        & $-1.16$ & $-2.11$    & $-1.26$     & $-3.62$ & $-3.02$ \\ \hline
\end{tabular}
\label{table:skewness_value}
\end{table*}

\Paragraph{Algorithm.}
Upon training job arrival, \sys{} first obtains an ordered set of layers, $\{L_1, L_2 \cdots, L_N\}$,
where $L_i$ is the $i$-th layer in the forward-pass layers.
It then retrieves the parameter size $P_i$ and output size $O_i$ of each layer $L_i$
and identifies the layer type (e.g., convolution layer) associated with $L_i$.
Using this information, \sys{} goes through the following steps.

(i) \textbf{Step 1:} It measures the layer parameter skewness $S$ of the CNN model to estimate what fraction of the model parameters are concentrated in the latter layers.
Formally, we measure the skewness using normality testing (with degree 3)~\cite{bai2005tests} represented as follows in our context:
\vspace{-0.7cm}
\begin{center}
\begin{equation} \label{eq:standardized moment} \tag{1} 
\frac{\frac{1}{n}\sum_{i=1}^{n}\left ( P_{i} - \bar{P} \right )^3}{\left ( \frac{1}{n}\sum_{i=1}^{n}\left ( P_{i} - \bar{P} \right )^2 \right )^{3/2}} \notag
\end{equation}
\end{center}  
\noindent
where $\bar{P}$ is the mean of all $P_i$ covering all layers in the model.

The result of Equation~\ref{eq:standardized moment} represents the skewness factor $S$ of the parameter size distribution across layers.
As a general rule of thumb, if $S$ < $0$, the distribution is left-skewed (i.e., data concentrated on the latter layers), and the absolute value of $S$ decides how highly the distribution is skewed.
Since CNN models under our study all show a large fraction of data used by the latter layers,
as Table~\ref{table:skewness_value} reports, the $S$ values of all 8 benchmarks used in our study are below zero.

(ii) \textbf{Step 2:} \sys{} filters out some models that violate a predefined criteria.
\sys{} steers a predefined threshold $K$ to decide how aggressively it enables layer placement.
For this, \sys{} compares the threshold $K$ with the output of Step 1, $S$,
and leverages a general rule of thumb that $-1$ < $S$ < $-0.5$ indicates a moderate level of skewness~\cite{normality_testing}.
Thus, by setting $K$ = $-0.5$, \sys{} Profiler decides to parallelize models as long as they exhibit at least moderate skewness.
In other instances, \sys{} Profiler could be very selective in applying model partitioning only to highly skewed models by setting $K$ to larger values.

(iii) \textbf{Step 3:} Next, \sys{} decides where to partition between the layers to be network-cost effective. 
This step requires \sys{} to scan all layers from $L_1$ to $L_N$, and estimate the network cost assuming that it splits the model with respect to the encountered layer.
\sys{} finds the layer that produces the smallest network cost in a single training step, calculated from the following equation:
\vspace{-0.7cm}
\begin{center}
\begin{equation*} \label{eq:netcost}
\arg\min_{1\leq i \leq N} \{O_i + \sum_{x=0}^{i}{P_x}\}
\end{equation*}
\end{center}

\noindent where $O_i$ is the output size of layer $L_i$ and 
$\sum_{x=0}^{i}{P_x}$ is the parameter size summed up to $L_i$ including all the preceding layers,
which need to be transferred at model aggregation.
Notice that the cost excludes constant factors such as the number of communication occurrences for a step.
\sys{} reports the minimum-cost layer as a position to split as long as it is not 
between compute-demand layers.

\subsection{Summary}
\sys{} improves distributed training of CNN using three insights: (i)
it colocates the memory-demand layers with PS in the same machine
to significantly reduce network communication, which used to be the main bottleneck;
(ii) it assigns more resources to workers that run the compute-intensive layers while assigning fewer GPUs to PS workers, providing an efficient way of using available compute resources;
and (iii) it profiles the CNN model using a well-defined cost model and guides partitioning when it turns out to be beneficial.

We believe that any machine learning framework can integrate \sys{} as long as
it exposes the model internal structure.
Moreover, our profiler and cost model are generally applicable to
other classes of neural networks if network communication is prohibitively expensive.
We also note that partitioning may need model-specific knowledge
such as the computational cost for the layers to be offloaded; e.g., for CNN, we avoid offloading the convolution layers.


\vspace{-0.2cm}
\section{Implementation}
\label{section:impl}
\vspace{-0.2cm}


%
We implement RALP in the TensorFlow framework. In TensorFlow, each layer is
organized as a data flow graph composed of nodes. 
TensorFlow provides \texttt{tf.device} as a means to place the data flow
graph of layers to particular servers. Unfortunately, \texttt{tf.device} can
only place the data flow graph within the \texttt{tf.device} scope. Hence,
only the compute node of the layers can be placed accordingly. As this does
not satisfy the needs of RALP, we take a lower level approach.

For RALP, we make changes to \texttt{device\_function()} in TensorFlow. This
is the function that sets the device in which the nodes of the data flow
graph are to be executed. 
Simply, we should modify \texttt{device\_function()} to place all nodes from the
layers determined through the skewness normality test in the parameter server.
However, there are some other matters that need to be considered.
First, there are some graph nodes that need to be excluded. Recall that the
loss value is calculated in between the forward and backward passes. Some of
these calculations make use of the memory-demand layer gradient values. Thus,
nodes that compute the loss value based on the memory-demand layer gradient
should not be placed on the parameter server. Finally, there are extra nodes
that need to be included in the parameter server, though this applies only
when there are multiple GPUs in a worker. When there are multiple GPUs, the
average value of the gradients resulting from the GPU is first calculated
before they are transmitted to the parameter server. The nodes that perform
these calculations are generated separately after the model graph is
constructed. Thus, we select the nodes that calculate the average of the
gradients of the compute-intensive layer and place them on the parameter
server.


Finally, to make use of RALP, we simply add a statement
``\texttt{tf.name\_scope(`RALP'):}'' to
the model configuration code to specify the \texttt{name\_scope}, and everything else is done automatically.

\section{Evaluation}
\label{section:evaluation}

In this section, we evaluate \sys{} on a number of benchmarks
to show that \sys{} achieves higher throughput over
the traditional PS architecture. We also show that \sys{} outperforms Horovod, 
which is the state-of-the-art in managing communication bandwidth usage for DDL training.
We then evaluate \sys{} for tasks that train on the dataset carrying out more
complex classification, which we expect to be more prevalent in the future.
Lastly, we evaluate the efficacy of \sys{} when multiple training jobs
run simultaneously in a cluster sharing network, validating the
significance of communication-aware layer placement on consolidated workloads.

\begin{figure*}[t]
\center
\centering
\includegraphics[width=18cm, clip]{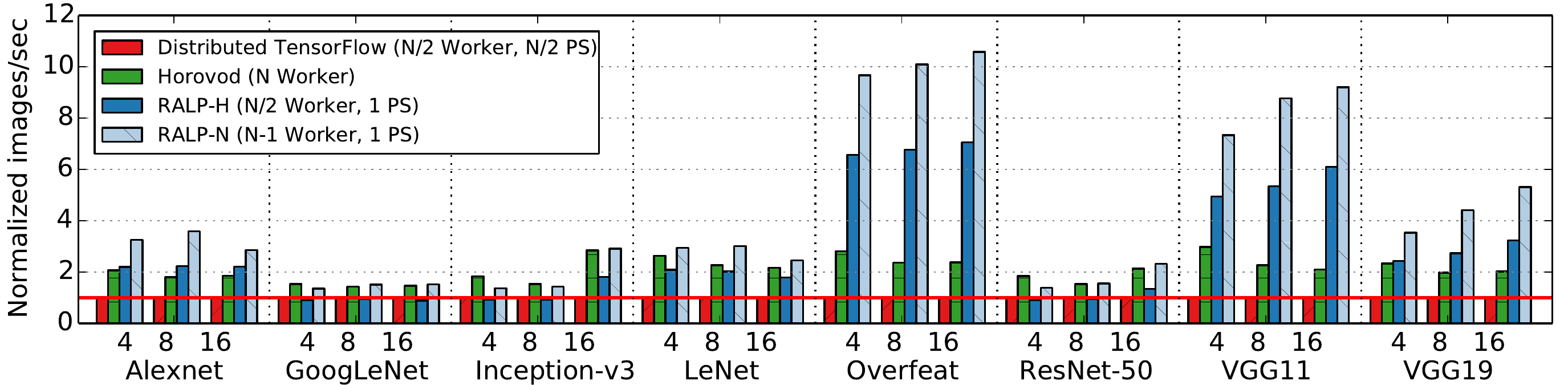}
\caption{Training performance with ImageNet-1K for Distributed TensorFlow (baseline), Horovod, and two configurations of RALP. The numbers on the $x$-axis represents number of GPUs = N.}
\label{figure:training performance}
\end{figure*}

\subsection{Experimental Setup}

\Paragraph{Testbed.}
Each machine has two 2.10~GHz 4-core Intel Xeon processors and 
64~GB of main memory, and 4 NVIDIA TITAN Xp 
GPUs each with 12~GB GPU memory.
We conduct experiments on a cluster of 8 machines, thus 32 GPUs in total.
The machines are connected via 56~Gbps RDMA (InfiniBand) network.

\Paragraph{Benchmarks and datasets.}
Each GPU in use runs either a worker and a parameter server (or PS). We vary the degree of
data parallelism by changing the number of workers, and additionally the number of PSes.


For evaluation we select the following 8 CNN benchmarks provided by TensorFlow~1.12~\cite{tf_benchmakrs}:
Alexnet~\cite{2012alexnet}, GoogLeNet~\cite{szegedy2015going}, Inception-v3~\cite{inception-v3_arxiv},
LeNet~\cite{lecun1998gradient}, Overfeat~\cite{2014overfeat}, ResNet-50~\cite{2016resnet},
VGG11~\cite{2015vgg}, and VGG19~\cite{2015vgg}.
We omit a few benchmarks that are variants of the selected benchmarks
since their trends were very similar with the ones under the same umbrella:
e.g., Inception-v3 and Inception-v4 exhibit similar trends in performance changes with \sys{}.
The batch size for each benchmark is set as the default value configured in TensorFlow~1.12.
Unless specified otherwise, in \sys{} we configure the threshold value to compare with skewness factor as -0.5
to enable layer placement for all benchmarks (see Table~\ref{table:skewness_value}) by default.

In all experiments, aside from the experiments in Section~\ref{section:evaluation_complexity_classification}, we use the ImageNet-1K~\cite{imagenet_dataset} dataset that
classifies images into 1,000 categories. 
We use the `ILSVRC 2017'~\cite{imagenet_2017} images as input data for training,
where there are 1,281,166 images in total.
For the experiments in Section~\ref{section:evaluation_complexity_classification}, we use the ImageNet-22K~\cite{imagenet_dataset} to evaluate \sys{} in the scenario where classification tasks become more complicated.
Due to the output size that increases with the number of image categories,
using ImageNet-22K needs to transfer more data over the network for model aggregation.

\Paragraph{Performance metric.}
As performance metric, we use throughput (images processed per second) across all workers at the training iteration. 
At the same time, we measure the data volume transferred between workers and PSes to show the effect \sys{} has on mitigating network traffic. 
We collect these metrics averaged over 100 iterations after warming up the system through the first few iterations.
We do not report convergence or accuracy of the trained model as there is no change in execution and thus, the accuracy remains the same.

\begin{figure*}[t!]
\vspace{-0.3cm}
\centering
\subfigure{
\includegraphics[width=0.235 \linewidth]{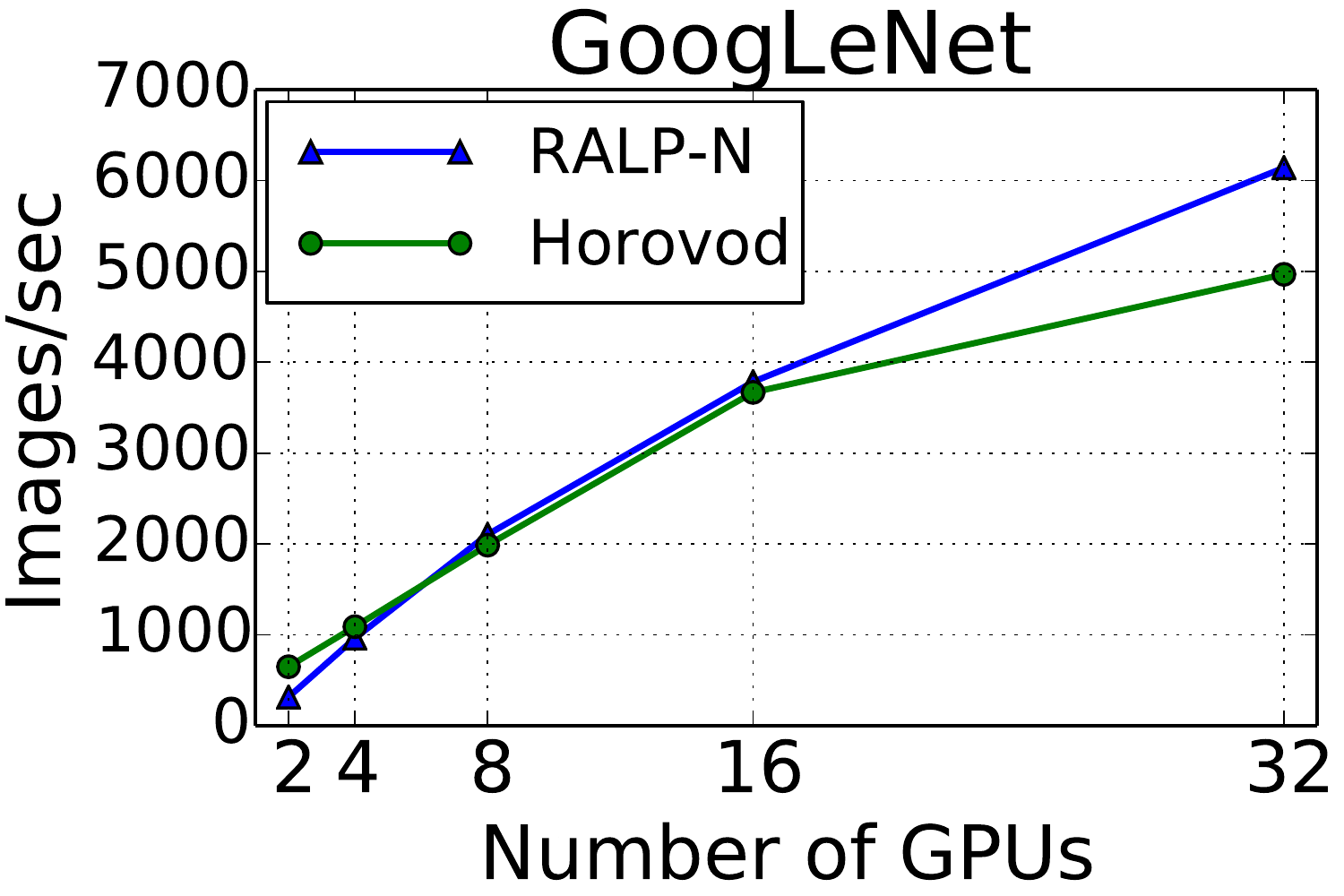}
}
\centering
\subfigure{
\includegraphics[width=0.235 \linewidth]{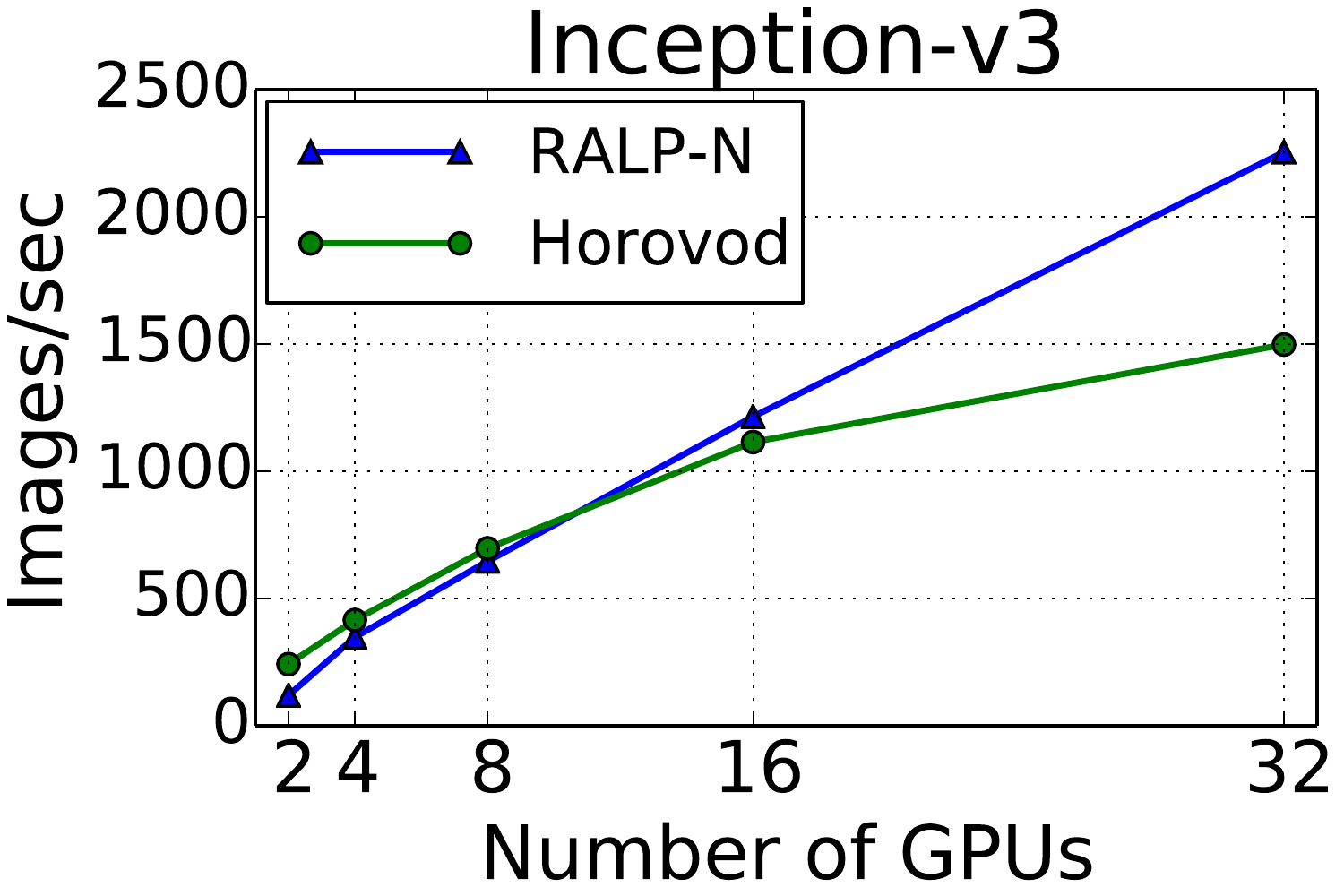}
}
\centering
\subfigure{
\includegraphics[width=0.235 \linewidth]{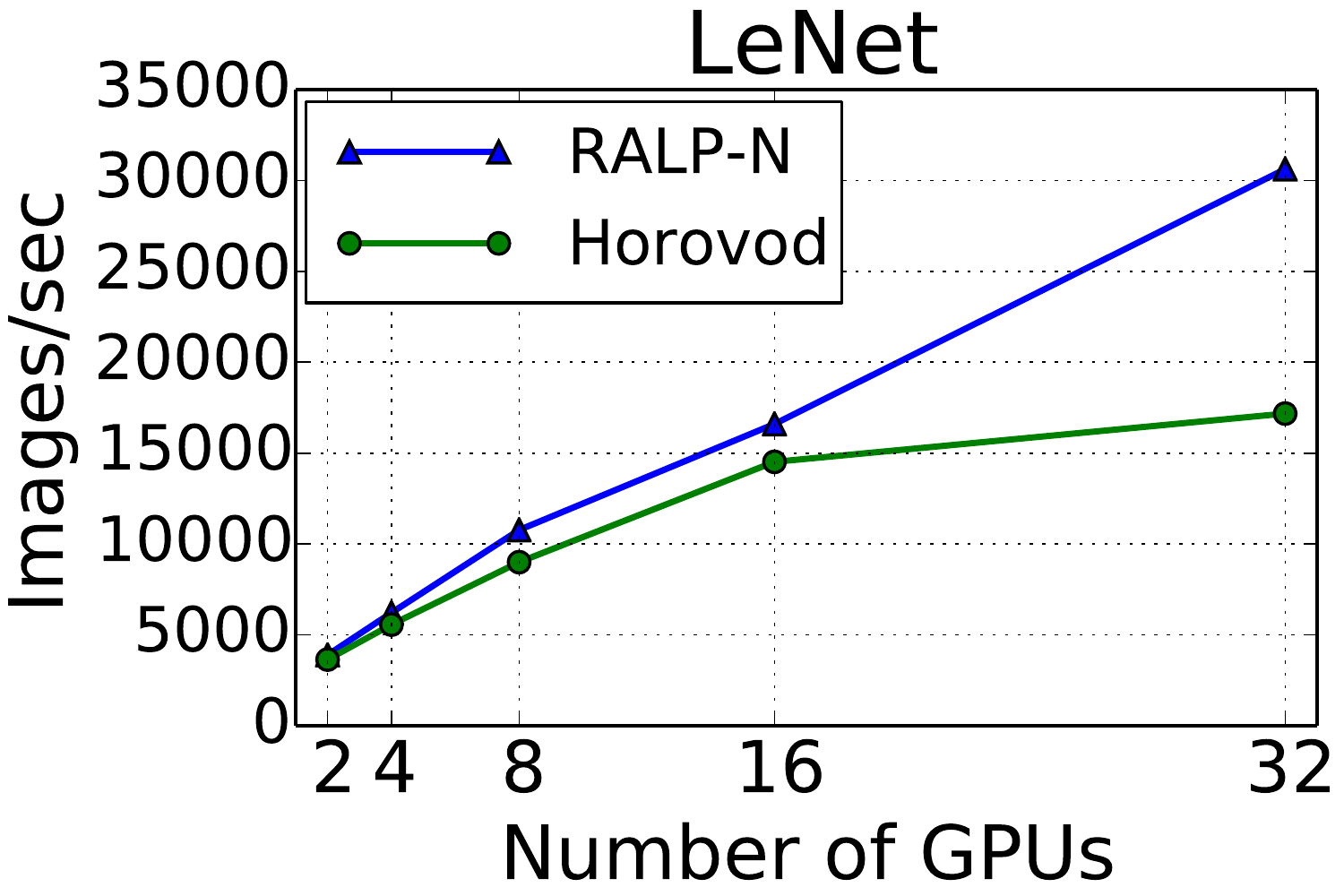}
}
\centering
\subfigure{
\includegraphics[width=0.235 \linewidth]{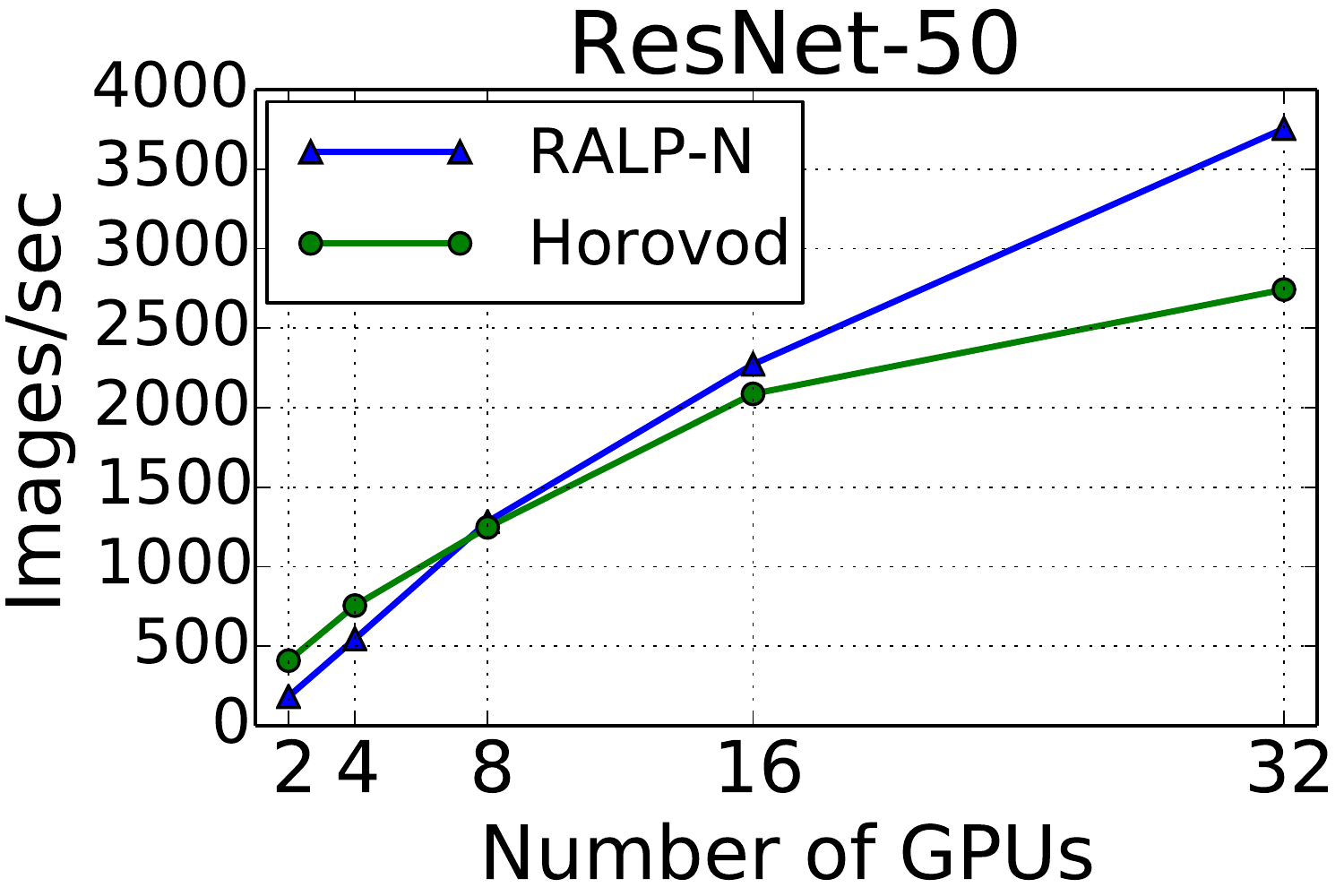}
}
\vspace{-0.1cm}
\caption{Comparison between \sys{}-N and Horovod while scaling up to 32 GPUs}
\label{figure:horovod performance}
\end{figure*}

\subsection{Single Model Training Performance}
\label{section:evaluation_single_model}

In this section, we look into the performance of \sys{} for isolated training on a variety of GPU assignment scenarios among workers and PSes.
We carry out experiments by assigning GPUs for a training job in machines fully distributed across the network; we evaluated packing training job's GPUs onto a smaller number of machines and observed similar trends.
The baseline for our comparison is distributed TensorFlow that has been optimized to scale in the PS architecture in TensorFlow~1.12~\cite{distributed_tensorflow}.
Also, to maintain good training speed in distributed TensorFlow, given $N$ GPUs we assign workers and PSes an equal number of GPUs (i.e., $N/2$ GPUs) as was done in previous work~\cite{peng2018optimus, nsdi_tiresias}.

For \sys{}, in comparison, we use a different configuration as the efficient handling of network cost allows us to run fewer PSes.
Thus, we evaluate \sys{} under two resource assignment scenarios:
(i) $N/2$ GPUs for workers and 1 GPU for PS (i.e., total $N/2+1$ GPUs), which we denote as \sys{}-H (for half, though not exactly, of the GPUs), aiming at saving resources,
and (ii) $N-1$ GPUs for workers and 1 GPU to PS (i.e., total $N$ GPUs), which we denote as \sys{}-N (for N GPUs), aiming at maximizing performance gain.
For comparison, we also show the performance of Horovod, the state-of-the-art that accelerates distributed training through mitigation of network usage~\cite{horovod}.
Horovod integrates a synchronization method called ring-reduce, which optimizes over all-reduce.
As this method operates without parameter servers, for Horovod, we use all the GPUs to run only the workers.

%
%

Figure~\ref{figure:training performance} shows throughput results, normalized to the baseline, using 4, 8, and 16 GPUs.
A few observations can be made from these results.
First, we find that \sys{}-H and \sys{}-N both outperform the baseline for all cases, with models such as Overfeat and VGG11 showing speedup of 5 to 10 times.
Thus, \sys{} can be used for either cost savings by using fewer GPUs (\sys{}-H) or more performance gains while using all assigned GPUs (\sys{}-N).
For a few benchmarks the performance of \sys{}-H is almost identical to the baseline as the benefit of saving network through layer offloading in \sys{} is largely offset by the cost to pay for transferring intermediate output data between workers and the PS.
Nonetheless, we see that \sys{}-H can save resource usage up to 7 out of
16 GPUs, and the saved resources can be used for extra capabilities,
e.g., improving capacity when consolidating training workloads.

\begin{table}[]
\centering
\caption{Transfer sizes of a 32-GPU training for one step.}
\begin{tabular}{l||c|c|c}
\hline
             & Model size & Horovod & \sys{} \\ \hline \hline
Alexnet      & 0.23~GB    & 3.23~GB & 0.47~GB \\
Inception-v3 & 0.09~GB    & 1.24~GB & 1.30~GB \\
VGG11        & 0.50~GB    & 6.93~GB & 0.64~GB \\ \hline
\end{tabular}
\label{table:horovod_size}
\end{table}

Second, we see that the performance of \sys{}-N is almost always better than Horovod.
For Alexnet, Overfeat, VGG11, and VGG19, the performance improvement is significant ranging from roughly 120\% to around 958\%.
On the other hand, for GoogLeNet, Inception-v3, LeNet, and ResNet-50, performance improvements are marginal, and there are even points where Horovod performs better than \sys{}-N.
However, once we increase the number of GPUs even further, we find that \sys{}-N starts to outperform Horovod by wider margins as shown in Figure~\ref{figure:horovod performance}.
The primary factor that leads to such superior performance with \sys{} relates to how much communication is saved. 
Formally, Horovod transfers data as much as $2\times S \times (W-1)$ for each training step, where S is the model size and W is the number of workers.
Table~\ref{table:horovod_size} reports the calculated numbers for three benchmarks for Horovod when using 32 GPUs (i.e., W$=$32).
Among the benchmarks, VGG11, which has the largest model size, has the biggest cost with 6.93~GB of data transferred at each step, which roughly translates to 8.97~GB/s of network transfer rate in our setup.
In contrast, \sys{}'s transfer volume is independent of the model size S. Instead, through offloading layers to the PS, it avoids network transfer for large-size layers that decide most of the model size. 
Consequently, \sys{} transfers much less data than Horovod as shown in Table~\ref{table:horovod_size}.
Our results are especially interesting as Horovod claims to scale well to the number of GPUs~\cite{horovod}.
Our results show that \sys{} scales even better, which leads to our last observation as follows.

Our final observation from Figure~\ref{figure:training performance} is that \sys{} provides a setting for higher parallelism.
As more GPUs are used to run more workers in \sys{}-N, we see a steady improvement in the relative throughput for most cases.
This leads to the conclusion that managing network communication is crucial in distributed training as the degree of parallelism becomes higher.
As a result, we can effectively steer \sys{} to trade off resource savings for
higher performance when idle GPUs are ample.

\Paragraph{Can \sys{} decide not to partition models with either no or minimal benefits?}
Yes. \sys{} can configure how aggressively it will perform layer placement
by simply steering its threshold value to compare with skewness factor.
For example, if the threshold is -1.5, \sys{} will trigger layer placement for Alexnet, Overfeat, VGG11, and VGG19 only,
which have skewness factors smaller than -1.5. 
Notice that these benchmarks show substantial
performance improvements as shown in Figure~\ref{figure:training performance}.

\begin{figure*}[t]
\center
\centering
\includegraphics[width=18cm, clip]{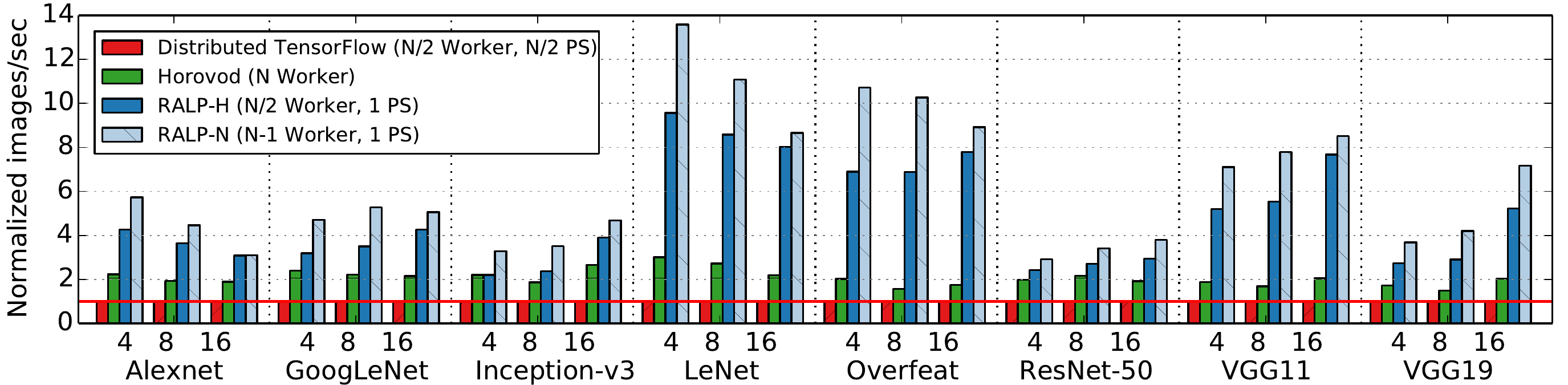}
\caption{Training performance with ImageNet-22K for Distributed TensorFlow (baseline), Horovod, and two configurations of RALP. The numbers on the $x$-axis represents number of GPUs = N.}
\label{figure:training performance 22K}
\end{figure*}

\subsection{Increased Complexity in Classification}
\label{section:evaluation_complexity_classification}
We now consider performance for a larger dataset.
For this, we perform experiments with the ImageNet-22K dataset, which is a dataset with 21,841 classes using the `ImageNet Fall 2011 release' image data. Since the TensorFlow CNN benchmark program does not provide standard models for
ImageNet-22K, we modify the number of classes of ImageNet-1K to accommodate
the 21,841 classes. 
Naturally, as the class classification is now 21,841, the
output of the last fully connected layer increases to 21,841. Thus, the last
fully connected layer has more gradients to calculate for the increased
number of weights and biases. Note that as the use of CNN models becomes more
prevalent, classification into more classes should be expected in the future.

Figure~\ref{figure:training performance 22K} shows the results presented in similar manner with Figure~\ref{figure:training performance}.
Overall, the trend in performance is similar to ImageNet-1K with the following two key differences. 
First, the improvements brought about by \sys{} is substantially higher compared to both the baseline and Horovod.
(Note the $y$-axis is higher.) 
Second, even \sys{}-H beats Horovod.
This means that even with slightly over half the resources, \sys{} is outperforming the state-of-the-art technique.

\subsection{Workload Consolidation}

%

\begin{table}[]
\centering
\caption{Workload consolidation images/sec}
\begin{tabular}{l||c|c|c}
\hline
             & TensorFlow & \sys{}  & Speed up \\ \hline\hline
VGG11 $\times$ 8    & 91.3           & 651.8 & 7.14$\times$     \\ \hline
VGG11 $\times$ 4    & 120.6          & 592.2 & 4.91$\times$     \\
ResNet-50 $\times$ 4 & 351.9          & 386.9 & 1.10$\times$     \\ \hline
ResNet-50 $\times$ 8 & 400.5          & 409.8 & 1.02$\times$     \\ \hline
\end{tabular}
\label{table:workload_consolidation}
\end{table}

In this section, we consider the effect of \sys{} under workload consolidation,
where we run multiple training jobs simultaneously in our shared cluster.
For this study, we revert back to the ImageNet-1K dataset.
We carry out evaluations under different sets of consolidated workloads as follows:
(i) when the workload is homogeneous and consists of jobs that all benefit from \sys{} the most
(i.e., \sys{} friendly jobs);
(ii) when the workload is homogeneous and consists of jobs that do not benefit much from \sys{}
(i.e., \sys{} unfriendly jobs);
and (iii) when the workload is mixed with \sys{} friendly and unfriendly jobs.
VGG11 and ResNet-50 are selected as the \sys{} friendly and unfriendly job, respectively.
Each job runs on four GPUs, where workers are assigned three GPUs and the PS runs on a single GPU. 
Thus, in our testbed, which has a total of 32 GPUs, we can consolidate up to 8 training jobs.
Note that the TensorFlow setting is different from those of Sections~\ref{section:evaluation_single_model} and~\ref{section:evaluation_complexity_classification} where the workers and PSes are set to equal numbers.
This is to compare TensorFlow and \sys{} under the same configuration such that the difference between the two can be clearly identified.

Table~\ref{table:workload_consolidation} summaries the results showing that \sys{}
achieves overall higher performance when the workload includes \sys{} friendly jobs.
The most benefit comes from when the workload consists of \sys{} friendly jobs only. In particular, when running eight VGG11 jobs in the cluster, the average speedup with \sys{} is as much as 7.14 times, 
where speedup is calculated as the throughput with \sys{} divided by the throughput with TensorFlow.
Moreover, we still observe the benefit when the workload is only partially \sys{} friendly, as can be seen in the second row in Table~\ref{table:workload_consolidation} where VGG11 and ResNet-50 exhibits $4.91 \times$ and $1.1 \times$ speedup, respectively.
Note that ResNet-50, the unfriendly job, performance improves by 10\% as well.
This indicates that under workload consolidation where network is actively shared, reduction in network communication results in aggregated benefits across all training jobs, both friendly and unfriendly, which is exactly what \sys{} brings about.
Lastly, running only ResNet-50 jobs results in very small improvements, which is expected since \sys{} has limited benefit on ResNet-50 training itself.

\begin{table}[]
\centering
\caption{Workload mixed consolidation images/sec}
\begin{tabular}{l||c c|c}
\hline
             & TensorFlow & \sys{}  & Speed up \\ \hline\hline
VGG11 $\times$ 4    &   -        & 605.2 & 5.02$\times$     \\
ResNet-50 $\times$ 4 & 384.3          & - & 1.09$\times$     \\ \hline
VGG11 $\times$ 4    & 122.9          & - & 1.02$\times$     \\ 
ResNet-50 $\times$ 4 &    -       & 352.1 & 1.00$\times$     \\ \hline
\end{tabular}
\label{table:workload_consolidation_2}
\end{table}

Now we consider consolidating models with an intermix of approaches, that is, jobs with TensorFlow and \sys{} running simultaneously.
Again, we make use of \sys{} friendly and unfriendly jobs, VGG11 and ResNet-50.
Table~\ref{table:workload_consolidation_2} shows how the jobs are mixed as well as the results.
In the first 2 rows, VGG11 and ResNet-50 are executed with \sys{} and TensorFlow, respectively, each of 4 jobs and each job using 4 GPUs, while in the bottom 2 rows, the jobs are flipped to use the other schemes.
The results reiterate the fact that the benefits from friendly jobs spill over even to unfriendly jobs resulting in aggregated benefits.
Specifically, in the first two rows, VGG11 throughput improves by $5.02\times$, where this speedup is calculated by dividing the RALP result in Table~\ref{table:workload_consolidation_2} (605.2) by the VGG11$\times 4$ TensorFlow result in Table~\ref{table:workload_consolidation} (120.6).
Interestingly, and similar to the mixed workload results in Table~\ref{table:workload_consolidation}, performance of the consolidated ResNet-50 improves by around 10\% even when executed with TensorFlow as shown in the second row.
In contrast, when VGG11 is run with TensorFlow, we see that performance of neither the friendly nor the unfriendly job improve as no network communication reduction is provided.

\section{Related Work}
\label{section:related}

Numerous studies have been conducted to improve deep
neural network performance. In this section, we discuss these
studies focusing on efforts to reduce communication overhead
and distribute workloads among system resources. We also discuss
other methods that have been proposed.

\Paragraph{Efficient distributed training.} 
Our work shares some similarities with Stanza~\cite{2018Stanza} in that both
exploit the fact that data transfer is mostly attributed to the
fully connected layers while incurring insignificant computation.
Similar to \sys{}, Stanza proposes to separate out the fully connected layers from other layers.
However, it does not discuss when to do this.
Our work showed that blindly applying layer splitting is
not always beneficial. 
Therefore, \sys{} proposes using \sys{} Profiler to enable splitting only when it is truly beneficial.
Horovod proposes an efficient communication method, namely, ring-reduce
to improve network communication performance~\cite{horovod}. 
We compared \sys{} with Horovod
extensively and found that \sys{} scales better because its network cost does not depend on the model size.
Other wrok such as Project Adam~\cite{chilimbi2014project} and Poseidon~\cite{2017poseidon}
decompose gradients in the fully connected layer to reduce
communication traffic.
\sys{} could make use of these techniques to run multiple PS workers efficiently.

Some studies demonstrate the effectiveness of distributed training on other aspects.
GeePS uses GPU memory as a cache for manipulating large scale training~\cite{cui2016geeps}. 
Parameter Hub proposes a software
design that provides a streamlined gradient processing pipeline~\cite{2018PHub}.
Awan et al. propose a pipelined chain design for the MPI\_Bcast
collective operation~\cite{2017Ammar}.
These studies are orthogonal to \sys{}.

\Paragraph{Model partitioning.} 
Our work is partly inspired by OWT, which proposes to  jointly use multiple parallelism schemes, especially for CNN using data parallelism on the convolutional layers and model parallelism on the fully-connected layers~\cite{2014OWT}.
Similar to \sys{}, OWT exploits the disproportionate computation-communication
characteristics that appear in the CNN model layers.
Jia et al. propose layer-wise parallelism, enabling each individual layer to use an independent parallelism policy~\cite{2018JiaExploring}.
The scope of these studies is primarily on the use of intra-machine resources.
Instead, our focus is on training on physically distributed machines, where
such cost models proposed in OWT~\cite{2014OWT} and by Jia et al.~\cite{2018JiaExploring} are not applicable. 
We thus proposed a cost model facilitated for model partitioning in a distributed setup and presented extensive experimental results in this study.

As the model size becomes larger, model training needs to be efficiently done in resource-constrained environments.
Strads shows that well-scheduled model parallelism using parameter convergence
can perform better than normal training in some common machine learning applications~\cite{kim2016strads}.
PipeDream partitions layers of model to place them on all available GPUs while achieving its dual goal of balancing computation and minimizing communication~\cite{2018PipeDream}. 
It reduces communication for large DNNs relative to data-parallel training.
Gpipe partitions a model across different GPUs and automatically splits a mini-batch of training examples into smaller batch sizes called micro-batches~\cite{2018GPipe}.
To split the model automatically, Gpipe uses a heuristic-based partitioning algorithm.

\Paragraph{Gradient optimization.}
One of the reasons for poor scalability in distributed training is network communication bottleneck.
In effect, the key method used to reduce network communication cost could be to reduce the size of data that is transferred.
There are two methods for gradient optimization, namely, gradient quantization and gradient sparsification.
Gradient quantization focuses on manipulating gradients to reduce the amount of data transfers, using
ternary value~\cite{2017terngrad}, random discrete values~\cite{2017qsgd}, and low bitwidth~\cite{zhou2016dorefa}. 
Gradient sparsification focuses on dropping some gradients out to reduce that amount, by 
discarding gradients randomly~\cite{wangni2018gradient} or the smallest gradients~\cite{2017AjiSpareSGD}.
These approaches are complementary our work, and \sys{} can be used to reduce network communication further even when these approaches are applied.

\section{Conclusion}
\label{section:conclusion}

In this paper, we proposed a resource-aware layer placement (RALP) scheme to
alleviate the network bottleneck and balance computation to improve performance for CNN distributed training. To achieve our
goal, we first went through a thorough analysis of CNN characteristics
considering the computational and communication needs of the layers that
comprise CNN. We find that due to their characteristics, the convolution and
pooling layers should be placed together in a separate worker server, while
the fully connected layer should be placed in the parameter server to
maximize performance. We provided a method to easily incorporate such
observations in the TensorFlow framework. Experimental evaluations making use
of this method showed that many CNN models are able to reap significant
performance improvements compared to the baseline TensorFlow framework as well as Horovod, a recently proposed state-of-the-art method.

{\footnotesize
\bibliographystyle{plain}
\bibliography{bibliography}}
　
\end{document}